\documentclass[11pt,a4paper]{article}
\usepackage[T1]{fontenc}
\usepackage[latin9]{inputenc}
\usepackage{amsmath}
\usepackage{amssymb}
\usepackage{esint}

\makeatletter

\pdfpageheight\paperheight
\pdfpagewidth\paperwidth

\pdfoutput=1 

\usepackage{jheppub}

\usepackage{etoolbox}
    
    \patchcmd{\maketitle}{\@fpheader}{}{}{}

\usepackage{amsfonts}

\usepackage{bbm}

\title{\boldmath Hypersymmetry bounds and three-dimensional higher-spin black holes}

\author[a,b]{Marc Henneaux,}
\author[b]{Alfredo P\'{e}rez,}
\author[a,b]{David Tempo,}
\author[b]{and Ricardo Troncoso}

\affiliation[a]{Universit\'{e} Libre de Bruxelles and International Solvay Institutes,
ULB Campus Plaine C.P.231, B-1050 Bruxelles, Belgium.}
\affiliation[b]{Centro de Estudios Cient\'{i}ficos (CECs), Av. Arturo Prat 514, Valdivia,
Chile.}

\emailAdd{henneaux@ulb.ac.be}
\emailAdd{aperez@cecs.cl}
\emailAdd{tempo@cecs.cl}
\emailAdd{troncoso@cecs.cl}

\preprint{CECS-PHY-15/01}

\abstract{We investigate the hypersymmetry bounds on the higher spin black hole parameters that follow from the asymptotic symmetry superalgebra in higher-spin anti-de Sitter gravity in three spacetime dimensions.  We consider anti-de Sitter hypergravity for which the analysis is most transparent.  This is a $osp(1\vert 4) \oplus osp(1\vert 4)$ Chern-Simons theory which contains, besides a spin-$2$ field, a spin-$4$ field and a spin-$5/2$ field.  The asymptotic symmetry superalgebra is then the direct sum of two-copies of the hypersymmetric extension $W_{(2,\frac52,4)}$ of $W_{(2,4)}$, which contains fermionic generators of conformal weight $5/2$ and bosonic generators of conformal weight $4$ in addition to the Virasoro generators.  Following standard methods, we derive bounds on the conserved charges from the anticommutator of the hypersymmetry generators.  The hypersymmetry bounds are nonlinear and are saturated by the hypersymmetric black holes,  which turn out to possess $1/4$-hypersymmetry and to be ``extreme'', where extremality can be defined in terms of the entropy:  extreme black holes are those that fulfill the extremality bounds beyond which the entropy ceases to be a real function of the black hole parameters. We also extend the analysis to other $sp(4)$-solitonic solutions which are maximally (hyper)symmetric.}

\makeatother

\begin{document}
\maketitle \flushbottom

\section{Introduction\label{sec:Introduction}}

Higher-spin anti-de Sitter gravity in three dimensions \cite{Blencowe,Vasiliev:1995dn,Bergshoeff:1989ns}
has attracted in recent years a great deal of interest. One result
that triggered that interest has been the derivation of its asymptotic
algebra, which is given by two copies of a nonlinear $W$-algebra,
the detailed structure of which depends on the spin content of the
model \cite{Henneaux-HS,Theisen-HS}. The nonlinear $W$-algebra extends
by generators of higher conformal weights the asymptotic Virasoro
algebra of pure AdS gravity \cite{Brown:1986nw}. It has the same
classical central charge.

Another remarkable feature of the theory is that it admits black holes
endowed with charges carried by the extra generators \cite{GK,AGKP,CM,Ammon:2012wc,HPTT,PTTreview,Bunster:2014mua}.
These generalize the BTZ black holes of pure gravity \cite{BTZ,BHTZ}.
Although the causal structure of these black holes is difficult to
define given that the spacetime geometry associated with the metric
is not invariant under the gauge transformations of the higher spin
gauge fields, one can study their thermodynamics by considering the
Euclidean continuation of the theory. One can then define without
ambiguity entropy, temperature and chemical potentials conjugate to
the higher spin charges. In fact, it is through the Euclidean formulation
that black holes are defined, as solutions of the Euclidean theory
with solid torus topology that admit a well-defined thermodynamics.

The black hole entropy can be expressed as a function of the (extensive)
global charges, i.e, the mass, angular momentum (``spin-$2$'' charges)
and higher spin charges. The entropy takes real values only when these
parameters vary within a subregion of parameter space. Outside that
region, the solution is not a black hole because it has no sensible
thermodynamics (complex entropy). The extreme black holes are defined
to be those that lie precisely on the boundary of the allowed region
in parameter space. In the pure gravity case, where there exists a
well-defined geometry, this definition coincides with the definition
in terms of causal structure and coincident horizons. Extreme black
holes are characterized by a different topology in the Euclidean formulation,
because the thermal cycles, which are contractible in the non-extreme
case where the solution has solid torus topology, cease to be so in
the extremal case.

Extreme black holes in four dimensions have exceptional properties
\cite{Gibbons:1982se,Gibbons:1982ih,Gibbons:1982fy,Aichelburg:1983ux,Tod:1983pm,Romans:1991nq,BTZ_Gauss_Bonnet,Gibbons-Kallosk,Teitelboim_action_and_entropy,Hawking:1994ii},
which, in supersymmetric theories, can often be related to the fact
that they possess supersymmetry and saturate Bogomol'nyi bounds. This
feature was shown to hold also for extreme BTZ black holes in \cite{Coussaert:1993jp}.
In the framework of higher spin theory, supersymmetry is naturally
enlarged into a symmetry which contains fermionic generators of half-integer
spins greater than $1/2$, called ``hypersymmetry generators''.
The asymptotic symmetry superalgebra of the corresponding three-dimensional
models contains then fermionic generators of half-integer conformal
weights greater than $3/2$ \cite{Henneaux:2012ny}. One may therefore
wonder whether hypersymmetry has consequences similar to supersymmetry
and, in particular, whether it implies also bounds on the conserved
charges of higher spin black holes.

The simplest context in which to investigate this question is hypergravity,
which contains just the hypersymmetry generators and no other fermionic
symmetry. Hypergravity was pioneered in four dimensions in \cite{Aragone:1979hx,Berends:1979wu,Aragone:1980rk},
where no-go theorems were established. In three dimensions, however,
the no-go theorems can be evaded \cite{Aragone:1983sz}. Three-dimensional
AdS hypergravity is a Chern-Simons theory with gauge superalgebra
given by $osp(1\vert4)\oplus osp(1\vert4)$, with $sl(2,\mathbb{R})$
principally embedded in $sp(4)$ (in each sector). It has been the
recent focus of study in \cite{Chen:2013oxa,Zinoviev:2014sza}. The
theory contains a spin-$2$ field, a spin-$5/2$ field and a spin-$4$
field and has $\mathcal{N}=(1,1)$ hypersymmetry (that is, the minimum
amount of hypersymmetry in each sector, i.e., 4+4 hypersymmetry parameters).
In the limit where the cosmological constant vanishes, the spin-$4$
field can be consistently decoupled, reproducing the theory of \cite{Aragone:1983sz}
(see also \cite{Fuentealba:2015jma}).

We construct the black hole solutions of the theory, by introducing
the chemical potentials in a manner that preserves manifestly the
asymptotic symmetry. Preserving the asymptotic symmetry is of course
a mandatory step if one wants to discuss its implications! We discuss
the black hole thermodynamical properties and derive in particular
the entropy as a function of the black hole conserved charges. This
enables one to identify the extreme black holes.

We then investigate the hypersymmetry properties of the black holes.
We show that the maximum amount of hypersymmetry that the black holes
can have is one quarter of hypersymmetry on each side. The hypersymmetric
black holes turn out to be extreme, even though not all extreme black
holes are hypersymmetric. Furthermore, the hypersymmetric black holes
saturate exactly the nonlinear bounds on the charges that follow from
the asymptotic algebra. Therefore, hypersymmetry enables one to derive
interesting bounds on the charges, just as supersymmetry.

It should be stressed that the bounds that emerge are nonlinear and
ought to be so if the hypersymmetry bounds are to match the extremality
bounds, given that the entropy is a nonlinear function of the charges.

We then turn to a different class of solutions, the so-called $sp(4)$-solitons,
which by construction are maximally (hyper)symmetric, i.e., they are
invariant under global $OSp(1|4)$ transformations. Interestingly,
the hypersymmetry bounds imply that the spin-4 charge of this class
of solutions is nonnegative. Furthermore, it is found that the spin-4
chemical potential $\mu$ plays an interesting role, since it can
be suitably fixed so that the configuration is able to retain up to
half of the hypersymmetries in the Euclidean continuation. Indeed,
if $\mu$ were absent, the hypersymmetries would be completely broken
for ``hot $sp(4)$-solitons''.

Our paper is organized as follows. In Section \ref{sec:Asymptotic},
we describe the model and show that its asymptotic symmetry superalgebra
is given by two copies of the (classical limit of the) $WB_{2}$-superalgebra,
also known as $W\left(2,5/2,4\right)$-superalgebra \cite{FigueroaO'Farrill:1991pb,Bellucci:1994xa}.
We then prove (Section \ref{sec:HS-Bounds}) that this superalgebra
implies, with unitarity, nonlinear bounds on the bosonic charges,
which are obtained by considering the anticommutators between the
spin-$\frac{5}{2}$ charges. Next, in Section \ref{sec:solutions},
we construct and analyse the black hole solutions and the $sp(4)$-solitons,
discussing in particular the allowed range of parameters. We show
that the black holes fulfill the hypersymmetry bounds but that only
a subset of solitons, which we explicitly determine, do. Section \ref{sec:HS}
is devoted to the study of the hypersymmetries of the solutions, i.e.,
to solving the Killing vector-spinor equation in the black hole and
soliton backgrounds. We also show there that the hypersymmetric black
holes are extremal. Section \ref{sec:Concluding-remarks} collects
our concluding remarks. The paper is completed by two appendices.
Appendix \ref{sec:MR} gives our conventions on $osp(1\vert4)$ and
provides the technical details of the derivation of the asymptotic
symmetries. Appendix \ref{sec:BHsKRAUS} examines black holes associated
with a different embedding of $sl(2)$ into $osp(1\vert4)$.

\section{Asymptotic structure of hypergravity\label{sec:Asymptotic} }

\setcounter{equation}{0}AdS hypergravity in three-dimensional spacetimes
with the minimum amount of hypersymmetry in each sector can be described
as an $osp(1\vert4)\oplus osp(1\vert4)$ Chern-Simons theory. The
(anti)commutation relations of $osp(1\vert4)$ are explicitly: 
\begin{align}
\left[L_{i},L_{j}\right] & =\left(i-j\right)L_{i+j}\thinspace,\nonumber \\
\left[L_{i},U_{m}\right] & =\left(3i-m\right)U_{i+m}\thinspace,\nonumber \\
\left[L_{i},\mathcal{S}_{p}\right] & =\left(\frac{3}{2}i-p\right)\mathcal{S}_{i+p}\thinspace,\nonumber \\
\left[U_{m},U_{n}\right] & =\frac{1}{2^{2}3}\left(m-n\right)\left(\left(m^{2}+n^{2}-4\right)\left(m^{2}+n^{2}-\frac{2}{3}mn-9\right)-\frac{2}{3}\left(mn-6\right)mn\right)L_{m+n}\nonumber \\
 & +\frac{1}{6}\left(m-n\right)\left(m^{2}-mn+n^{2}-7\right)U_{m+n}\thinspace,\label{eq:WedgeAlgebra}\\
\left[U_{m},\mathcal{S}_{p}\right] & =\frac{1}{2^{3}3}\left(2m^{3}-8m^{2}p+20mp^{2}+82p-23m-40p^{3}\right)\mathcal{S}_{i+p}\thinspace,\nonumber \\
\left\{ \mathcal{S}_{p},\mathcal{S}_{q}\right\}  & =U_{p+q}+\frac{1}{2^{2}3}\left(6p^{2}-8pq+6q^{2}-9\right)L_{p+q}\thinspace.\nonumber 
\end{align}
Here $L_{i}$, with $i=0,\pm1$, stand for the spin-2 generators that
span the $sl\left(2,\mathbb{R}\right)$ subalgebra, while $U_{m}$
and ${\cal S}_{p}$, with $m=0,\pm1,\pm2,\pm3$ and $p=\pm\frac{1}{2},\pm\frac{3}{2}$,
correspond to the spin-4 and fermionic spin-$\frac{5}{2}$ generators,
respectively. Here, the word ``spin'' refers to the conformal weight
of the corresponding asymptotic generators in the conformal algebra
at infinity, see below.

The dynamics derives from the difference of two Chern-Simons actions,
$I=I_{CS}\left[A^{+}\right]-I_{CS}\left[A^{-}\right]$, with 
\begin{equation}
I_{CS}\left[A\right]=\frac{k_{4}}{4\pi}\int str\left[AdA+\frac{2}{3}A^{3}\right]\thinspace,\label{eq:ICS}
\end{equation}
where the level, $k_{4}=k/10$, is expressed in terms of the Newton
constant and the AdS radius according to $k=\ell/4G$. In eq.(\ref{eq:ICS})
$str\left[\cdots\right]$ stands for the supertrace of the fundamental
($5\times5$) matrix representation described in Appendix \ref{sec:MR}
and the gauge fields $A^{\pm}$ correspond to the two independent
copies of $osp(1\vert4)$.

As explained in \cite{CHvD}, by virtue of a suitable choice of gauge
group elements, $g_{\pm}=g_{\pm}\left(r\right)$, the radial dependence
of the asymptotic form of the connections can be completely gauged
away, so that $A^{\pm}=g_{\pm}^{-1}a^{\pm}g_{\pm}+g_{\pm}^{-1}dg_{\pm}$,
with 
\begin{equation}
a^{\pm}=a_{\varphi}^{\pm}\left(t,\varphi\right)d\varphi+a_{t}^{\pm}\left(t,\varphi\right)dt\thinspace.
\end{equation}
Then, following the lines of \cite{Henneaux-HS,Theisen-HS,HASugra,Henneaux:2012ny},
the asymptotic behaviour of the dynamical fields at a fixed time slice
$t=t_{0}$, is assumed to be such that deviations with respect to
the reference background go along the highest weight generators, i.e.,
\begin{align}
a_{\varphi}^{\pm} & =L_{\pm1}-\frac{2\pi}{k}\mathcal{L}^{\pm}\left(t,\varphi\right)L_{\mp1}+\frac{\pi}{5k}\mathcal{U}^{\pm}\left(t,\varphi\right)U_{\mp3}-\frac{2\pi}{k}\psi^{\pm}\left(t,\varphi\right)\mathcal{S}_{\mp\frac{3}{2}}\thinspace.\label{eq:atheta}
\end{align}
These fall-off conditions are maintained under a restricted set of
gauge transformations, $\delta a^{\pm}=d\Omega^{\pm}+\left[a^{\pm},\Omega^{\pm}\right]$,
where each of the Lie-algebra-valued parameters 
\[
\Omega^{\pm}=\Omega^{\pm}\left[\epsilon_{\pm},\chi_{\pm},\vartheta_{\pm}\right]\thinspace,
\]
depend on two bosonic and one Grassmann-valued functions of $t$ and
$\varphi$, given by $\epsilon_{\pm},\chi_{\pm}$, and $\vartheta_{\pm}$,
respectively. The explicit form of $\Omega^{\pm}$ is given in Appendix
\ref{sec:MR}. The fields $\mathcal{L}^{\pm}$, $\mathcal{U}^{\pm}$,
and $\psi^{\pm}$ transform in a precise way under the gauge transformations
generated by $\Omega^{\pm}$, which is also given in Appendix \ref{sec:MR}.

In order to preserve the asymptotic symmetries under evolution in
time, the asymptotic form of the Lagrange multiplier has to be of
the form 
\begin{equation}
a_{t}^{\pm}=\pm\Omega^{\pm}\left[\xi_{\pm},\mu_{\pm},\varrho_{\pm}\right]\thinspace,\label{eq:at}
\end{equation}
where $\Omega^{\pm}$ is defined in (\ref{eq:Omegamn}), and the ``chemical
potentials'' $\xi_{\pm},\mu_{\pm},\varrho_{\pm}$ are assumed to
be arbitrary functions of $t,\varphi$ that are fixed at the boundary.
It is worth pointing out that the asymptotic form of $a_{t}$ is preserved
under the asymptotic symmetries provided the field equations are fulfilled
in the asymptotic region, and the parameters $\epsilon,\chi,\vartheta$
that describe the asymptotic symmetries satisfy suitable ``deformed
chirality conditions\textquotedblright{} given by differential equations
of first order in time (see \cite{HPTT,Bunster:2014mua}).

Following the canonical approach \cite{Regge:1974zd}, the asymptotic
structure described above allows to readily find the explicit form
of the generators associated to the asymptotic symmetries \cite{Henneaux:2012ny},
which reads 
\begin{equation}
{\cal Q}^{\pm}\left[\epsilon_{\pm},\chi_{\pm},\vartheta_{\pm}\right]=-\int d\varphi\left(\epsilon_{\pm}{\cal L}^{\pm}+\chi_{\pm}{\cal U}^{\pm}-i\vartheta_{\pm}\psi^{\pm}\right)\thinspace.
\end{equation}
Since the Poisson brackets fulfill $\left[{\cal Q}\left[\eta_{1}\right],{\cal Q}\left[\eta_{2}\right]\right]_{PB}=-\delta_{\eta_{1}}{\cal Q}\left[\eta_{2}\right]$,
the algebra of the canonical generators can be easily found by virtue
of the transformation law of the fields in eqs. (\ref{eq:DeltaFields0})
and (\ref{eq:DeltaFields}). In Fourier modes, $X=\frac{1}{2\pi}\sum_{m}X_{m}e^{im\varphi}$,
the algebra is given by 
\begin{align}
i\left[L_{m},L_{n}\right]_{PB} & =\left(m-n\right)L_{m+n}+\frac{k}{2}m^{3}\delta_{m+n}^{0}\thinspace,\nonumber \\
i\left[L_{m},U_{n}\right]_{PB} & =\left(3m-n\right)U_{m+n}\thinspace,\nonumber \\
i\left[L_{m},\psi_{n}\right]_{PB} & =\left(\frac{3}{2}m-n\right)\psi_{m+n}\thinspace,\nonumber \\
i\left[U_{m},U_{n}\right]_{PB} & =\frac{1}{2^{2}3^{2}}\left(m-n\right)\left(3m^{4}-2m^{3}n+4m^{2}n^{2}-2mn^{3}+3n^{4}\right)L_{m+n}\nonumber \\
 & +\frac{1}{6}\left(m-n\right)\left(m^{2}-mn+n^{2}\right){\cal U}_{m+n}-\frac{2^{3}3\pi}{k}\left(m-n\right)\Lambda_{m+n}^{\left(6\right)}\nonumber \\
 & -\frac{7^{2}\pi}{3^{2}k}(m-n)\left(m^{2}+4mn+n^{2}\right)\Lambda_{m+n}^{\left(4\right)}+\frac{k}{2^{3}3^{2}}m^{7}\delta_{m+n}^{0}\thinspace,\label{eq:AlgebraModes}\\
i\left[U_{m},\psi_{n}\right]_{PB} & =\frac{1}{2^{2}3}\left(m^{3}-4m^{2}n+10mn^{2}-20n^{3}\right)\psi_{m+n}-\frac{23\pi}{3k}i\Lambda_{m+n}^{\left(11/2\right)}\nonumber \\
 & +\frac{\pi}{3k}\left(23m-82n\right)\Lambda_{m+n}^{\left(9/2\right)}\thinspace,\nonumber \\
i\left[\psi_{m},\psi_{n}\right]_{PB} & =U_{m+n}+\frac{1}{2}\left(m^{2}-\frac{4}{3}mn+n^{2}\right)L_{m+n}+\frac{3\pi}{k}\Lambda_{m+n}^{\left(4\right)}+\frac{k}{6}m^{4}\delta_{m+n}^{0}\thinspace,\nonumber 
\end{align}
where $\Lambda_{m}^{\left(l\right)}$ stand for the expansion of the
nonlinear terms in (\ref{eq:NonLinearterms}), while fermionic modes
are labeled by integers or half-integers in the case of periodic or
antiperiodic boundary conditions, respectively. The modes are assumed
to fulfill the following reality conditions: $\left(L_{m}\right)^{*}=L_{-m}$,
$\left(U_{m}\right)^{*}=U_{-m}$, $\left(\psi_{m}\right)^{*}=\psi_{-m}$
so that the functions ${\cal L}^{\pm}$, ${\cal U}^{\pm}$ and $\psi^{\pm}$
are real.

The fermionic functions $\psi^{\pm}$ are naturally antiperiodic when
the spatial sections are homeomorphic to the plane. However, for the
punctured plane topology, different spin structures are possible and
the fermionic functions $\psi^{\pm}$ can be periodic or antiperiodic.

In the case of antiperiodic boundary conditions, the wedge algebra
of (\ref{eq:AlgebraModes}) reduces {for each copy to $osp\left(1|4\right)$.
Indeed, dropping the nonlinear terms, the algebra (\ref{eq:AlgebraModes})
reproduces the one in (\ref{eq:WedgeAlgebra}) once the modes are
restricted according to $\left|n\right|<s$, where $s$ is the conformal
weight of the generators, provided the zero mode of $L_{n}$ is shifted
as $L_{0}\rightarrow L_{0}+\frac{k}{4\pi}$.

The asymptotic symmetry algebra (\ref{eq:AlgebraModes}) corresponds
to the classical limit of $WB_{2}$, also known as $W\left(2,5/2,4\right)$
\cite{FigueroaO'Farrill:1991pb,Bellucci:1994xa}. The spin-$\frac{5}{2}$
fermionic generators $\psi_{m}$ are the ``hypersymmetry'' generators.

\section{Hypersymmetry bounds from asymptotic algebra \label{sec:HS-Bounds}}

\subsection{Derivation of the bounds}

It is natural to wonder whether hypersymmetry has similar consequences
as supersymmetry. In particular, it is of interest to investigate
whether hypersymmetry also implies the existence of bounds for the
bosonic conserved charges. In order to pursue this task, we consider
bosonic configurations carrying global charges with only zero modes,
given by $L_{0}=2\pi{\cal L}$ and $U_{0}=2\pi{\cal U}$, for each
copy (``rest frame'' \cite{Bunster:2014mua}).

It turns out that hypersymmetry bounds do exist, and that these searched-for
hypersymmetry bounds can be derived from unitarity by following the
same semi-classical considerations as in the case of supergravity
\cite{Deser-Teitelboim,Teitelboim_Susy,Witten-positivity,Abott-Deser,Hull:1983ap,Teitelboim:1984kf}.
It should be noted in that respect that the quantum $W\left(2,5/2,4\right)$-superalgebra,
with the unitarity conditions $L_{m}^{\dagger}=L_{-m}$, $U_{m}^{\dagger}=U_{-m}$,
$\psi_{m}^{\dagger}=\psi_{-m}$ implied by the classical reality conditions,
admits arbitrarily large values of the central charge \cite{FigueroaO'Farrill:1991pb,Bellucci:1994xa}.
This allows one to take the classical limit. In the sequel, we shall
drop terms that are subdominant in that limit.

To derive the hypersymmetry bounds, one starts from the anticommutators
of the hypersymmetry generators in (\ref{eq:AlgebraModes}) with $m=-n=p\geq0$.
In the rest frame, these are found to reduce to 
\begin{align}
\left(2\pi\right)^{-1}\left(\hat{\psi}_{p}\hat{\psi}_{-p}+\hat{\psi}_{-p}\hat{\psi}_{p}\right) & =\hat{{\cal U}}+\frac{5}{3}p^{2}\hat{\mathcal{L}}+\frac{3\pi}{k}\hat{\mathcal{L}^{2}}+\frac{k}{12\pi}p^{4}\thinspace.
\end{align}
Now, the hermitian operator $\hat{\psi}_{p}\hat{\psi}_{-p}+\hat{\psi}_{-p}\hat{\psi}_{p}$
is positive definite. This implies, in the classical limit, that the
global charges fulfill the bounds 
\begin{equation}
B_{p}\equiv{\cal U}+\frac{5}{3}p^{2}\mathcal{L}+\frac{3\pi}{k}\mathcal{L}^{2}+\frac{k}{12\pi}p^{4}\geq0\thinspace.\label{bounds0}
\end{equation}
These bounds are manifestly nonlinear.

\subsection{Strongest bound}

In the periodic case, the bound $B_{0}\geq0$ for $p=0$ reads 
\begin{equation}
B_{0}\equiv{\cal U}+\frac{3\pi}{k}\mathcal{L}^{2}\geq0\thinspace,
\end{equation}
and one has 
\begin{equation}
B_{p}=B_{0}+\frac{5}{3}p^{2}\mathcal{L}+\frac{k}{12\pi}p^{4}\,.
\end{equation}
This shows that when $\mathcal{L}\geq0$, the bound for $p=0$ is
the strongest one in the sense that $B_{0}\geq0$ implies $B_{p}>0$.
In the anti-periodic case, one can also write $B_{p}=B_{\frac{1}{2}}+A_{p}$
where $A_{p}$ is manifestly positive when $\mathcal{L}\geq0$, but
this observation is less useful than in the periodic case because
$\mathcal{L}$ is in general not positive for the anti-periodic physically
interesting solutions.

\subsection{Expression in terms of eigenvalues}

It is convenient to express the bounds as 
\begin{align}
\left(p^{2}+\lambda_{\left[+\right]}^{2}\right)\left(p^{2}+\lambda_{\left[-\right]}^{2}\right) & \geq0\thinspace,\label{eq:BoundLambdamn}
\end{align}
with 
\begin{align}
\lambda_{\left[\pm\right]}^{2} & =\frac{10\pi}{k}\left(\mathcal{L}\pm\frac{4}{5}\sqrt{\mathcal{L}^{2}-\frac{3k}{16\pi}\mathcal{U}}\right)\;.\label{eq:Lambda2mn}
\end{align}
This is because the $\pm\lambda_{\left[\pm\right]}$ precisely coincide
with the (non-zero) eigenvalues of the dynamical gauge field $a_{\varphi}$,
given by 
\begin{equation}
\lambda_{\left[\pm\right]}^{2}=\frac{1}{4}\left(str\left[a_{\varphi}^{2}\right]\pm\sqrt{4str\left[a_{\varphi}^{4}\right]-str\left[a_{\varphi}^{2}\right]^{2}}\right)\thinspace.\label{eq:CharPolyEig-aphi}
\end{equation}
Note that the eigenvalues of an $sp(4,\mathbb{R})$ matrix come in
opposite pairs $\pm\lambda$, and that if $\lambda$ is an eigenvalue,
so is $\lambda^{*}$. Since the holonomies of the dynamical gauge
field (\ref{eq:atheta}) along a circle 
\begin{equation}
{\cal H}_{\varphi}={\cal P}e^{\int a_{\varphi}d\varphi}=e^{2\pi a_{\varphi}}\thinspace,\label{eq:Holo-phi-1}
\end{equation}
are completely characterized by the eigenvalues of $a_{\varphi}$,
the bounds (\ref{eq:BoundLambdamn}) can be viewed, for each integer
or half-integer $p$ (depending on the periodicity conditions) as
bounds on the holonomies.

\subsection{The case of real eigenvalues}

An interesting and important case corresponds to real eigenvalues
$\lambda_{\left[\pm\right]}$, because the bounds (\ref{eq:BoundLambdamn})
are then automatically fulfilled. The condition that the eigenvalues
are real implies that ${\cal L}\geq0$ since ${\cal L}$ is, up to
a positive numerical factor, the sum of the squares of the eigenvalues.
Furthermore, one has 
\begin{equation}
-{\cal L}^{2}\leq\frac{k}{3\pi}{\cal U}\leq\frac{2^{4}}{3^{2}}{\cal L}^{2}\thinspace.\label{eq:BoundsGen}
\end{equation}
The second inequality expresses that the argument under the square
root in (\ref{eq:Lambda2mn}) is a positive real number, so that the
squares of the eigenvalues are real. The first inequality expresses
that $\lambda_{\left[-\right]}^{2}\geq0$. When $\lambda_{\left[-\right]}^{2}=0$,
the bound (\ref{bounds0}) (or (\ref{eq:BoundLambdamn})) with $p=0$
(``strongest bound''), available for periodic boundary conditions,
is saturated. The other bounds ($p>0$) are fulfilled, but never saturated.

It should be stressed that the status of the two inequalities in (\ref{eq:BoundsGen})
is rather different. Only the first one does saturate one of the bounds
in (\ref{eq:BoundLambdamn}) and can be related as we shall discuss
to hypersymmetry, corresponding to states annihilated by $\hat{\psi}_{0}$.
So, while the first inequality is a consequence of hypersymmetry with
$p=0$ and holds whether the eigenvalues are real or not, the other
inequality arises from the extra requirement that the eigenvalues
be real, which is not an a priori necessity and is indeed not fulfilled
by some of the solutions.

\section{Bosonic solutions}

\label{sec:solutions}

We now explore the set of regular bosonic solutions of the theory
that fulfill the bounds (\ref{eq:BoundLambdamn}), including the cases
that saturate them, which are expected to possess globally-defined
``Killing vector-spinors''.

\subsection{General considerations}

The searched for solutions only carry zero-mode bosonic charges that
fulfill the asymptotic fall-off in (\ref{eq:atheta}), and hence,
they are globally described by gauge fields of the form 
\begin{align}
a_{\varphi}^{\pm} & =L_{\pm1}-\frac{2\pi}{k}\mathcal{L}^{\pm}L_{\mp1}+\frac{\pi}{5k}\mathcal{U}^{\pm}U_{\mp3}\thinspace,\label{eq:athetaBH}
\end{align}
where $\mathcal{L}^{\pm}$ and $\mathcal{U}^{\pm}$ stand for the
global charges. Since the solutions do not carry fermionic charges,
their corresponding ``chemical potentials'' $\varrho_{\pm}$ can
be consistently set to vanish; and for the sake of simplicity, we
assume hereafter that the remaining (bosonic) ones $\xi_{\pm},\mu_{\pm}$,
which are held fixed at the boundary in the grand canonical ensemble,
are constant. Hence, according to (\ref{eq:at}), the Lagrange multipliers,
given by the time component $a_{t}^{\pm}$ of the gauge fields, are
given by 
\begin{align}
a_{t}^{\pm} & =\xi_{\pm}a_{\varphi}^{\pm}+\mu_{\pm}\left[-U_{\pm3}-\frac{6\pi}{k}\mathcal{U}^{\pm}L_{\mp1}+\frac{6\pi}{k}\mathcal{L}^{\pm}U_{\pm1}\right.\nonumber \\
 & \left.-\frac{\pi}{k}\left(\mathcal{U}^{\pm}+\frac{12\pi}{k}\left(\mathcal{L}^{\pm}\right)^{2}\right)U_{\mp1}+\frac{22}{15}\frac{\pi^{2}}{k^{2}}\left(\mathcal{U}^{\pm}+\frac{60\pi}{11k}\left(\mathcal{L}^{\pm}\right)^{2}\right)\mathcal{L}^{\pm}U_{\mp3}\right]\thinspace.\label{eq:at-BH}
\end{align}

Regularity of the solution then requires the holonomy along any contractible
cycle ${\cal C}$ to be trivial. In the case of $osp\left(1|4\right)$
this condition reads 
\begin{equation}
{\cal H}_{{\cal C}}={\cal P}e^{\int_{{\cal C}}a}=\Gamma^{\pm}=\left(\begin{array}{cc}
\pm\mathbbm{1}_{4\times4} & 0\\
0 & 1
\end{array}\right)\thinspace,\label{eq:TrivalHol-C}
\end{equation}
where the plus or the minus signs correspond to the different spin
structures, i.e., fermions that fulfill periodic or antiperiodic boundary
conditions, respectively. Note that $\Gamma^{-}$ anticommutes with
the fermionic generators, i.e.,$\left\{ \Gamma^{-},{\cal S}_{p}\right\} =0$,
as it should.

Depending on the solutions under consideration, the contractible circles
along which to impose triviality may correspond to either Euclidean
time or the angular coordinate. The corresponding configurations describe
black holes or solitonic-like solutions, respectively. The analysis
of these classes of solutions, including whether they fulfill or saturate
the bounds in (\ref{eq:BoundLambdamn}) is discussed in what follows.

\subsection{Black holes with spin-4 charges\label{sec:HS-BH}}

\subsubsection{Euclidean continuation}

Whenever massless higher spin fields are included, the metric does
not define an invariant geometry since it transforms under the higher
spin gauge symmetries. This makes the definition of the black holes
in terms of horizons associated with the metric causal structure unavailable.
Black holes can still be defined, however, through the Euclidean continuation,
as solutions admitting a well-defined thermodynamics (real temperature,
entropy and thermodynamic parameters) \cite{GK,AGKP,Bunster:2014mua}.
This approach is equivalent to the metric one when both are at one's
disposal, but appears to be the only one for higher spin gravity.

As shown in \cite{Carlip:1994gc}, the Euclidean black hole manifold
has the topology of the solid torus for pure gravity. This remains
true for higher spin gravity \cite{GK,AGKP,Bunster:2014mua}. Furthermore,
the Euclidean continuation is defined by letting the gauge fields
take complex values in such a way that $A=A^{+}$ and $A^{\dagger}=-A^{-}$.
Therefore, the relationship between Euclidean fields and chemical
potentials with their Lorentzian counterparts reads 
\begin{align}
\mathcal{L} & =\mathcal{L}^{+}\;,\;\mathcal{U}=\mathcal{U}^{+}\;;\;\mathcal{L}^{\ast}=\mathcal{L}^{-}\;,\;\mathcal{U}^{\ast}=\mathcal{U}^{-}\thinspace,\\
\xi & =\xi_{+}\;,\;\mu=\mu_{+}\;;\;\xi^{\ast}=\xi_{-}\;,\;\mu^{\ast}=\mu_{-}\thinspace,
\end{align}
and hence, by virtue of (\ref{eq:atheta}) and (\ref{eq:at}), the
Euclidean solution is given by 
\begin{align}
a & =\left(L_{1}-\frac{2\pi}{k}\mathcal{L}L_{-1}+\frac{\pi}{5k}\mathcal{U}U_{-3}\right)d\varphi-\left\{ i\xi\left(L_{1}-\frac{2\pi}{k}\mathcal{L}L_{-1}+\frac{\pi}{5k}\mathcal{U}U_{-3}\right)-i\mu\left[U_{3}-\frac{6\pi}{k}\mathcal{L}U_{1}\right.\right.\nonumber \\
 & \left.-\frac{22}{15}\frac{\pi^{2}}{k^{2}}\left(\mathcal{U}+\frac{60\pi}{11k}\mathcal{L}^{2}\right)\mathcal{L}U_{-3}+\frac{\pi}{k}\left(\mathcal{U}+\frac{12\pi}{k}\mathcal{L}^{2}\right)U_{-1}+\frac{6\pi}{k}\mathcal{U}L_{-1}\right\} d\tau\thinspace,\label{eq:aEucl}
\end{align}
where $\tau=it$.

It is worth emphasizing that since the chemical potentials are already
incorporated along the temporal components of the connection, the
analysis can be performed for a fixed range of the coordinates. Thus,
we assume henceforth that $0<\tau\le1$, and $0<\varphi\le2\pi$.

\subsubsection{Black hole solutions}

On the solid torus, the temporal (``thermal'') circles corresponding
to Euclidean time are contractible. These are the only contractible
cycles along which the regularity condition is not automatically satisfied
and must therefore be imposed. Hence the regularity condition of the
gauge field (\ref{eq:TrivalHol-C}) reads for black hole solutions
\begin{equation}
{\cal H}_{\tau}=e^{a_{\tau}}=\Gamma^{\pm}\thinspace,\label{eq:Holo-tau}
\end{equation}
which implies that the eigenvalues of $a_{\tau}$, that are obtained
by making the replacement $a_{\varphi}\rightarrow a_{\tau}$ in eq.
(\ref{eq:CharPolyEig-aphi}) have to be given by $\pm2\pi im$ and
$\pm2\pi in$, with $m$, $n$ being integers or half-integers for
$\Gamma^{+}=\mathbbm{1}$ or $\Gamma^{-}$, respectively. The regularity
condition then reduces to 
\begin{align}
str\left[a_{\tau}^{4}\right] & =32\pi^{4}\left(m{}^{4}+n{}^{4}\right)\thinspace,\\
str\left[a_{\tau}^{2}\right] & =-8\pi^{2}\left(m{}^{2}+n{}^{2}\right)\thinspace,
\end{align}
so that once evaluated in the Euclidean solution (\ref{eq:aEucl}),
it reads 
\begin{align*}
\frac{1}{2}\left(\frac{k}{\pi}\xi\right)^{4}\left(41\mathcal{L}^{2}-\frac{3k}{\pi}\mathcal{U}\right)+\frac{3k}{\pi}\left(2^{4}\mu\right)^{3}\xi\left[\mathcal{L}^{5}+\frac{k}{48\pi}\left(41\mathcal{L}^{2}-\frac{5^{2}k}{2^{3}3\pi}\mathcal{U}\right)\mathcal{L}\mathcal{U}\right]\\
-3\mu\left(\frac{4k}{\pi}\xi\right)^{3}\left(\mathcal{L}^{2}-\frac{17k}{24\pi}\mathcal{U}\right)\mathcal{L}+3^{3}\left(\frac{2^{3}k}{\pi}\mu\xi\right)^{2}\left[\mathcal{L}^{4}-\frac{3k}{2^{4}\pi}\left(\mathcal{L}^{2}-\frac{5^{2}k}{3^{3}\pi}\mathcal{U}\right)\mathcal{U}\right]\\
+\left(2^{3}\mu\right)^{4}\left\{ \frac{41}{2}\mathcal{L}^{6}+\frac{7k}{12\pi}\left[2^{3}3^{2}\mathcal{U}\mathcal{L}^{4}+\frac{5^{2}k}{2^{5}\pi}\left(\frac{23}{3}\mathcal{L}^{2}-\frac{5^{2}k}{7\pi}\mathcal{U}\right)\mathcal{U}^{2}\right]\right\} -\frac{k^{6}}{\pi^{2}}\left(m{}^{4}+n{}^{4}\right) & =0\thinspace,\\
2^{6}\mu^{2}\mathcal{L}^{3}+\frac{28k}{3\pi}\left(\mu^{2}\mathcal{U}+\frac{3k}{28\pi}\xi^{2}\right)\mathcal{L}+\frac{2^{2}k^{2}}{\pi^{2}}\mu\xi\mathcal{U}-\frac{k^{3}}{5\pi}\left(m{}^{2}+n{}^{2}\right) & =0\thinspace.
\end{align*}
Generically, this condition allows to express the chemical potentials
in terms of the global charges, or equivalently, in terms of the eigenvalues
of $a_{\varphi}$, given by (\ref{eq:CharPolyEig-aphi}), and therefore
\begin{align}
\xi & =\frac{\pi}{5^{2}}\left[\frac{9m\lambda_{\left[-\right]}^{3}-41\left(m\lambda_{\left[+\right]}-n\lambda_{\left[-\right]}\right)\lambda_{\left[-\right]}\lambda_{\left[+\right]}-9n\lambda_{\left[+\right]}^{3}}{\left(\lambda_{\left[-\right]}^{2}-\lambda_{\left[+\right]}^{2}\right)\lambda_{\left[-\right]}\lambda_{\left[+\right]}}\right]\thinspace,\nonumber \\
\mu & =\frac{6\pi}{5}\left[\frac{m\lambda_{\left[-\right]}-n\lambda_{\left[+\right]}}{\left(\lambda_{\left[-\right]}^{2}-\lambda_{\left[+\right]}^{2}\right)\lambda_{\left[-\right]}\lambda_{\left[+\right]}}\right]\thinspace.\label{eq:ximuRC}
\end{align}
Note that for generic values of $m,n$, the holonomy around the thermal
circle becomes nontrivial in the cases that fulfill $\left(\lambda_{\left[-\right]}^{2}-\lambda_{\left[+\right]}^{2}\right)\lambda_{\left[-\right]}\lambda_{\left[+\right]}=0$,
and consequently, the chemical potentials turn out to be arbitrary.
This corresponds to the extremal cases, with zero temperature $T$
($\xi\rightarrow\infty$), for which the topology of the Euclidean
manifold changes\footnote{Remarkably, the extremal solutions can be ``regularized'' for special
values of $m,n$, so that the topology remains the one of a solid
torus. This occurs in the following cases: $m=0$ ($\lambda_{\left[+\right]}=0$),
$n=0$ ($\lambda_{\left[-\right]}=0$), $m=n$ ($\lambda_{\left[+\right]}=\lambda_{\left[-\right]}$),
and $m=-n$ ($\lambda_{\left[+\right]}=-\lambda_{\left[-\right]}$),
where the brackets stand for the cases that are regularized. None
of these cases corresponds to the BTZ branch.}.

The branch that it is connected with the BTZ black hole, for which
${\cal U}=0$ and ${\cal L}>0$, and hence $\lambda_{\left[+\right]}=3\lambda_{\left[-\right]}=3\sqrt{\frac{2\pi}{k}{\cal L}}$,
corresponds to $n=\frac{1}{2}$ and $m=\frac{3}{2}$.

\subsubsection{Entropy}

Following the canonical approach, the correct expression for the black
hole entropy was found in \cite{PTT1,PTT2}, and further developed
in \cite{deBoer:2013gz} and \cite{Bunster:2014mua}. In our conventions,
which agree with the ones in \cite{Bunster:2014mua}, the black hole
entropy is given by 
\begin{align}
S & =-2k_{4}\text{Im}\left(\text{str}\left[a_{\tau}a_{\varphi}\right]\right)_{\text{on-shell}}\;,\label{eq:entropy}
\end{align}
which for the solution (\ref{eq:aEucl}), is readily evaluated to
be 
\begin{align}
S & =8\pi\mbox{Re}\left[\xi\mathcal{L}+2\mu\mathcal{U}\right]_{\text{on-shell}}\ .
\end{align}
The regularity conditions in (\ref{eq:ximuRC}) then imply 
\begin{equation}
S=\frac{4\pi k}{5}\mbox{Re}\left(m\lambda_{\left[+\right]}+n\lambda_{\left[-\right]}\right)\thinspace,
\end{equation}
which, by virtue of (\ref{eq:Lambda2mn}), allows one to express the
black hole entropy in terms of the global charges ${\cal L}$, ${\cal U}$.
For the branch that is connected to the BTZ black hole ($n=\frac{1}{2},\;m=\frac{3}{2}$),
the entropy is given by 
\begin{equation}
S=\frac{2\pi k}{5}\mbox{Re}\left(3\lambda_{\left[+\right]}+\lambda_{\left[-\right]}\right)\thinspace.\label{eq:Slambda}
\end{equation}
In terms of the global charges, the entropy reads 
\begin{equation}
S=4\pi\sqrt{2\pi k\mathcal{L}}\text{\thinspace Re}\left[\frac{1}{2\sqrt{5}}\left(\sqrt{1-\frac{4}{5}\sqrt{1-\frac{3k\mathcal{U}}{16\pi\mathcal{L}^{2}}}}+3\sqrt{1+\frac{4}{5}\sqrt{1-\frac{3k\mathcal{U}}{16\pi\mathcal{L}^{2}}}}\right)\right]\thinspace.
\end{equation}
Therefore, in the Lorentzian continuation the entropy acquires the
following form: 
\begin{align}
S & =\pi\sqrt{\frac{2}{5}\pi k}\left[\sqrt{\mathcal{L}^{+}}\left(\sqrt{1-\frac{4}{5}\sqrt{1-\frac{3k\mathcal{U}^{+}}{16\pi\left(\mathcal{L}^{+}\right)^{2}}}}+3\sqrt{1+\frac{4}{5}\sqrt{1-\frac{3k\mathcal{U}^{+}}{16\pi\left(\mathcal{L}^{+}\right)^{2}}}}\right)\right.\nonumber \\
 & +\left.\sqrt{\mathcal{L}^{-}}\left(\sqrt{1-\frac{4}{5}\sqrt{1-\frac{3k\mathcal{U}^{-}}{16\pi\left(\mathcal{L}^{-}\right)^{2}}}}+3\sqrt{1+\frac{4}{5}\sqrt{1-\frac{3k\mathcal{U}^{-}}{16\pi\left(\mathcal{L}^{-}\right)^{2}}}}\right)\right]\thinspace.
\end{align}

In order to have a sensible thermodynamics, the entropy must be a
real quantity. This is so provided the eigenvalues $\lambda_{\left[\pm\right]}$
are real. Thus, one must impose ${\cal L}^{\pm}\geq0$, and, as explained
in section \ref{sec:HS-Bounds}, 
\begin{equation}
-\left({\cal L}^{\pm}\right)^{2}\leq\frac{k}{3\pi}{\cal U}^{\pm}\leq\frac{2^{4}}{3^{2}}\left({\cal L}^{\pm}\right)^{2}\thinspace.\label{eq:BoundLor}
\end{equation}

\subsubsection{Extremal black holes}

Extremal black holes are defined to be black holes, the parameters
of which lie at the frontier of the allowed range. Thus, we have two
types of extremal black holes, those for which 
\begin{equation}
-\left({\cal L}^{\pm}\right)^{2}=\frac{k}{3\pi}{\cal U}^{\pm}\thinspace,\label{eq:BoundLor22}
\end{equation}
and those for which 
\begin{equation}
\frac{k}{3\pi}{\cal U}^{\pm}=\frac{2^{4}}{3^{2}}\left({\cal L}^{\pm}\right)^{2}\thinspace.\label{eq:BoundLor33}
\end{equation}
Note that when ${\cal L}^{\pm}=0$, the spin 4 charge vanishes, ${\cal U}^{\pm}=0$.
This is the zero mass BTZ black hole.

The first type of extremal black holes (\ref{eq:BoundLor22}) saturates
the hypersymmetry bounds (\ref{bounds0}). And conversely, the black
holes that saturate the hypersymmetry bound (\ref{bounds0}) are extremal.
This is because the bounds (\ref{bounds0}) and (\ref{eq:BoundLor22})
are identical. For this family, the spin-4 charge is non-positive.

The other type (\ref{eq:BoundLor33}) of extremal black holes do not
saturate the hypersymmetry bound. Note, however, that they do fulfill
these unitarity bounds since the eigenvalues are real. We do not have
yet a direct physical interpretation of this other extremality bound
(\ref{eq:BoundLor33}).

It is worth highlighting that the black hole entropy is sensitive
to the sign of the spin-4 charges ${\cal U}^{\pm}$. Indeed, for a
fixed value of the spin-2 charges, the allowed range of positive spin-4
charges turns out to be wider, according to the bounds in (\ref{eq:BoundLor}),
than that of the negative ones.

\subsection{Spin 4 solitonic-like solutions\label{sec:HS-SOL}}

When the contractible circle along which the regularity condition
is not automatic corresponds to the angular coordinate, one finds
solitonic-like solutions. These solutions are defined in the Lorentzian
sector and are the analogs of the conical defects and surpluses investigated
in \cite{Castro:2011iw,Camp-Conical-defects,Camp-Freden,Li-Lin-Wang,Raey1}.

The regularity condition (\ref{eq:TrivalHol-C}) now reads 
\begin{equation}
{\cal H}_{\varphi}=e^{2\pi a_{\varphi}}=\Gamma^{\pm}\thinspace,\label{eq:Holo-phi}
\end{equation}
for both copies of the gauge field. It implies that the eigenvalues
of $a_{\varphi}$ in eq. (\ref{eq:CharPolyEig-aphi}) are given by
$\pm im$ and $\pm in$, with $m,n\ne0$, and $m^{2}\text{\ensuremath{\neq}}n^{2}$,
where $m$, $n$ stand for integers or half-integers for $\Gamma^{+}$or
$\Gamma^{-}$, respectively. To avoid multiple counting of solutions,
we shall impose without loss of generality $m>n>0$, and we also examine
for completeness both the anti-periodic and periodic boundary conditions.
When regularity holds, the global charges are consequently quantized
according to 
\begin{align}
\mathcal{L} & =-\frac{k}{20\pi}\left(m^{2}+n^{2}\right)\thinspace\thinspace\thinspace;\thinspace\thinspace\thinspace\mathcal{U}=\frac{k}{2^{4}5^{2}3\pi}\left(m^{2}-9n^{2}\right)\left(n^{2}-9m^{2}\right)\thinspace,\label{eq:LUmn}
\end{align}
expressions which are symmetric under $m\leftrightarrow n$.

As explained in section \ref{sec:HS-Bounds}, the bosonic charges
have to fulfill the hypersymmetry bounds in eq. (\ref{eq:BoundLambdamn}),
which become 
\begin{equation}
\left(p^{2}-m^{2}\right)\left(p^{2}-n^{2}\right)\geq0\thinspace.
\end{equation}
These bounds must hold for {\em all} $p$'s. Now, if $p\geq m$,
the bounds are satisfied since both $p^{2}-m^{2}$ and $p^{2}-n^{2}$
are then $\geq0$. If $p<m$, the bound will be satisfied only if
$p\leq n$, and this will occur for all $p<m$ if and only if $n=m-1$.
We thus see that the subset of solitonic-like solutions for which
the bounds hold are characterized by 
\begin{equation}
m=n+1\thinspace.\label{eq:mncondSoliton}
\end{equation}
The solutions which do not fulfill this condition are not compatible
with unitarity (in the large $c$ limit we are considering), which,
as we have seen, underlies the hypersymmetry bounds.

We shall call the solitonic-like solutions that fulfill (\ref{eq:mncondSoliton})
the ``allowed class''. This allowed class of solitonic-like solutions
is labelled by a single integer $n$, and the global charges are given
by 
\begin{align}
\mathcal{L} & =-\frac{k}{20\pi}\left(2n^{2}+2n+1\right)\leq-\frac{k}{8\pi}\thinspace,\label{eq:LUabN}
\end{align}
with 
\begin{align}
\mathcal{U} & =\frac{16\pi}{3k}\left(\mathcal{L}+\frac{k}{32\pi}\right)\left(\mathcal{L}+\frac{k}{8\pi}\right)\geq0\thinspace,
\end{align}
so that negative spin-4 charges become excluded.

For any solution in the allowed class, the hypersymmetry bounds are
saturated for two values of $p$, given by $p=n$ and $p=m=n+1$.
It is also interesting to note that the upper bound for the charges
${\cal L}$ in (\ref{eq:LUabN}) is attained for $n=\frac{1}{2}$,
available in the case of antiperiodic boundary conditions. This configuration
has ${\cal L}^{\pm}=-\frac{k}{8\pi}$ and saturates also the lower
bound ${\cal U}^{\pm}=0$ for the other charge ${\cal U}^{\pm}$.
It is just AdS$_{3}$. In the case of periodic boundary conditions,
the extremal values for the charges cannot be reached and the solutions
are necessarily endowed with spin-4 charges, whose lowest values,
given by $\mathcal{U}^{\pm}=\frac{7k}{2^{4}3\pi}$ are attained for
$n=1$. This solution has also largest values of the spin-2 charges,
given by ${\cal L}^{\pm}=-\frac{k}{4\pi}$.

\section{Solutions with unbroken hypersymmetry\label{sec:HS}}

\subsection{Killing vector-spinor equation}

We explore in this section the hypersymmetry properties of the bosonic
solutions described above, given by (\ref{eq:aEucl}). That is, we
look for configurations that are globally invariant under hypersymmetry
transformations. The invariance condition reads 
\begin{equation}
\delta a=d\theta+\left[a,\theta\right]=0\thinspace,\label{eq:KVS-eq}
\end{equation}
where the gauge parameter $\theta$ is fermionic, i.e., $\theta=\theta^{p}{\cal S}_{p}$
for both copies. Eq. (\ref{eq:KVS-eq}) is the ``Killing vector-spinor
equation''.

Since the connection is flat, it can locally be expressed as $a=g^{-1}dg$.
It follows that the general solution of the Killing vector-spinor
equation (\ref{eq:KVS-eq}) is given by $\theta=g^{-1}\theta_{0}g$,
where $\theta_{0}$ is a constant chosen in such a way that the product
$g^{-1}\theta_{0}g$ is globally well defined (even though $g$ is
generically not). This shows that the maximum number of hypersymmetries
is equal to four (number of components of $\theta_{0}$) on each side.
There will be less hypersymmetries in general because of the constraints
on $\theta_{0}$.

In order to obtain the explicit form of the Killing vector-spinors
$\theta$, we follow a shortcut that circumvents finding explicitly
the $5\times5$ matrix group element $g$. By definition, global symmetries
are particular asymptotic symmetries. Hence the Killing vector-spinors
are of the form $\theta=\Omega\left[0,0,\vartheta\right]$, with $\Omega$
given by (\ref{eq:Omegamn}). In the case of the plus copy ($a_{\varphi}=a_{\varphi}^{+}$)
on which we shall again focus our attention for definiteness, $\theta$
is given by 
\begin{align}
\theta & =-\vartheta{\cal S}_{\frac{3}{2}}+\vartheta^{\prime}{\cal S}_{\frac{1}{2}}-\frac{1}{2}\left(\vartheta^{\prime\prime}-\frac{6\pi}{k}\mathcal{L}\vartheta\right){\cal S}_{-\frac{1}{2}}+\frac{1}{6}\left(\vartheta^{\prime\prime}-\frac{14\pi}{k}\mathcal{L}\vartheta\right)^{\prime}{\cal S}_{-\frac{3}{2}}\thinspace.\label{eq:theta}
\end{align}
where $\vartheta=\vartheta(t,\varphi)$. This function must now be
determined so that $\delta a_{\varphi}=0$ strictly, and not just
asymptotically.

Preserving the form of the dynamical fields, $\delta a_{\varphi}=0$,
reduces to the requirement that $\delta{\cal L}=\delta{\cal U}=\delta\psi=0$.
Since the spin-$5/2$ field $\psi$ is equal to zero for the above
solutions, the variation of the bosonic fields identically vanishes.
The variation of the fermionic ones yields the following condition
\begin{equation}
\delta\psi=-\left(\mathcal{U}+\frac{3\pi}{k}\mathcal{L}^{2}\right)\vartheta+\frac{5}{3}\mathcal{L}\vartheta^{\prime\prime}-\frac{k}{12\pi}\vartheta^{\prime\prime\prime\prime}=0\thinspace.\label{eq:KVS-eqphi}
\end{equation}
Analogously, preserving the form of the Lagrange multipliers, $\delta a_{t}=0$,
gives 
\begin{equation}
\dot{\vartheta}=\left[\left(\xi+\frac{82\pi}{3k}\mu\mathcal{L}\right)\vartheta-\frac{5}{3}\mu\vartheta^{\prime\prime}\right]^{\prime}\thinspace.\label{eq:KVS-eqT}
\end{equation}
The equations (\ref{eq:KVS-eqphi}), (\ref{eq:KVS-eqT}) can be trivially
integrated. To that end, it is useful to express them in terms of
the eigenvalues of $a_{\varphi}$, so that they read 
\begin{align}
 & \vartheta^{\prime\prime\prime\prime}-\left(\lambda_{\left[+\right]}^{2}+\lambda_{\left[-\right]}^{2}\right)\vartheta^{\prime\prime}+\lambda_{\left[+\right]}^{2}\lambda_{\left[-\right]}^{2}\vartheta=0\thinspace,\label{eq:KVS-phi}\\
\dot{\vartheta} & -\left[\left[\xi+\frac{41}{30}\mu\left(\lambda_{\left[+\right]}^{2}+\lambda_{\left[-\right]}^{2}\right)\right]\vartheta-\frac{5}{3}\mu\vartheta^{\prime\prime}\right]^{\prime}=0\thinspace.\label{eq:KVS-T}
\end{align}
In the case of the minus copy ($a_{\varphi}=a_{\varphi}^{-}$), the
Killing vector-spinor equations correspond to (\ref{eq:KVS-phi}),
(\ref{eq:KVS-T}), with $t\rightarrow-t$.

In the case of a generic bosonic solution, eq. (\ref{eq:KVS-phi})
integrates as 
\begin{equation}
\vartheta={\cal A}_{1}e^{\lambda_{\left[+\right]}\varphi}+{\cal A}_{2}e^{-\lambda_{\left[+\right]}\varphi}+{\cal A}_{3}e^{\lambda_{\left[-\right]}\varphi}+{\cal A}_{4}e^{-\lambda_{\left[-\right]}\varphi}\thinspace,\label{eq:varthetaG}
\end{equation}
were the function ${\cal A}_{I}={\cal A}_{I}\left(t\right)$ are restricted
by (\ref{eq:KVS-T}). We successively analyze these restrictions for
the black hole solutions and the soliton-like solutions.

\subsection{Black holes with spin-4 charges}

As discussed in section \ref{sec:HS-BH}, the eigenvalues are real
for a generic black hole solution. Accordingly, the vector-spinor
$\vartheta$ in (\ref{eq:varthetaG}) clearly cannot fulfill neither
periodic nor antiperiodic boundary conditions in the generic case.
Therefore, a generic black hole solution breaks all the hypersymmetries.
This is in agreement with the fact that they fulfill, but do not saturate,
the hypersymmetry bounds (\ref{eq:BoundLambdamn}). This situation
covers also the second type of extremal black holes with $\lambda_{\left[+\right]}=\lambda_{\left[-\right]}\neq0$
(${\cal U}=16\pi{\cal L}^{2}/(3k)$), for which, in spite of the fact
that equation (\ref{eq:KVS-phi}) integrates in a different way given
the degeneracy of the eigenvalues, 
\begin{equation}
\vartheta={\cal A}_{1}e^{\lambda_{\left[+\right]}\varphi}+{\cal A}_{2}e^{-\lambda_{\left[+\right]}\varphi}+{\cal A}_{3}\varphi e^{\lambda_{\left[+\right]}\varphi}+{\cal A}_{4}\varphi e^{-\lambda_{\left[+\right]}\varphi}\thinspace,\label{eq:varthetaG-1}
\end{equation}
the vector-spinor also fails to fulfill the periodicity conditions
unless the ${\cal A}_{I}$'s all vanish.

For extremal black holes of the first type, i.e., with $\lambda_{\left[-\right]}=0$
(${\cal U}=-3\pi{\cal L}^{2}/k$), the situation is different. The
solution of equation (\ref{eq:KVS-phi}) is given by (the eigenvalue
$0$ is degenerate) 
\begin{equation}
\vartheta={\cal A}_{1}e^{\lambda_{\left[+\right]}\varphi}+{\cal A}_{2}e^{-\lambda_{\left[+\right]}\varphi}+{\cal A}_{3}\varphi+{\cal A}_{4}\thinspace.\label{eq:varthetaG-1-1}
\end{equation}
This expression fulfills periodic boundary conditions only if ${\cal A}_{1}={\cal A}_{2}={\cal A}_{3}=0$.
The remaining equation (\ref{eq:KVS-T}) then reduces to $\dot{\vartheta}=0$,
and hence, this class of extremal black holes possesses a single constant
Killing vector-spinor $\vartheta$. This result goes hand in hand
with the fact that the hypersymmetry bound (\ref{eq:BoundLambdamn})
with $p=0$ is saturated. In the limiting case $\lambda_{\left[+\right]}=\lambda_{\left[-\right]}=0$
(${\cal L}={\cal U}=0$), one arrives to the same conclusion. Indeed,
equation (\ref{eq:KVS-phi}) now becomes $\vartheta^{\prime\prime\prime\prime}=0$,
which gives 
\begin{equation}
\vartheta={\cal A}_{3}\varphi^{3}+{\cal A}_{2}\varphi^{2}+{\cal A}_{1}\varphi+{\cal A}_{0}\thinspace,\label{eq:varthetaG-1-1-1}
\end{equation}
so that periodicity implies that $\vartheta={\cal A}_{0}\left(t\right)$,
while equation (\ref{eq:KVS-T}) again reads $\dot{\vartheta}=0$,
implying that $\vartheta$ is a constant.

In summary, black holes admit at most one quarter of the hypersymmetries
on each copy. Moreover, hypersymmetric black holes saturate the strongest
bound that follows from the asymptotic symmetry algebra, and they
are extreme. Nevertheless, not all extreme black holes are hypersymmetric.

\subsection{Spin-4 solitonic-like solutions}

It is easy to verify that for the solitonic-like configurations endowed
with spin-4 charges of section \ref{sec:HS-SOL}, the regularity conditions
imply that the solutions are invariant under global $OSp\left(1\vert4\right)$
transformations. Thence, they are maximally (hyper)symmetric.

The explicit expression of the Killing vector-spinors is given by
eq. (\ref{eq:varthetaG}), with $\lambda_{\left[+\right]}^{2}=-m^{2}$
and $\lambda_{\left[-\right]}^{2}=-n^{2}$, i.e., $\vartheta={\cal A}_{\left[+\right]}e^{im\varphi}+{\cal A}_{\left[+\right]}^{*}e^{-im\varphi}+{\cal A}_{\left[-\right]}e^{in\varphi}+{\cal A}_{\left[-\right]}^{*}e^{-in\varphi}\thinspace$,
where the (half) integers $m,n$ fulfill $m>n>0$. The form of the
functions ${\cal A}_{\left[\pm\right]}={\cal A}_{\left[\pm\right]}\left(t\right)$
is then fixed by eq. (\ref{eq:KVS-T}), and hence the Killing vector-spinors
are given by 
\begin{align}
\vartheta & ={\cal B}_{\left[+\right]}e^{im\left(\varphi+\omega_{m,n}t\right)}+{\cal B}_{\left[+\right]}^{*}e^{-im\left(\varphi+\omega_{m,n}t\right)}+{\cal B}_{\left[-\right]}e^{in\left(\varphi+\omega_{n,m}t\right)}+{\cal B}_{\left[-\right]}^{*}e^{-in\left(\varphi+\omega_{n,m}t\right)}\thinspace,\label{eq:vartheta-Sol1}
\end{align}
where 
\[
\omega{}_{m,n}=\xi-\frac{1}{30}\left(41m^{2}-9n^{2}\right)\mu\thinspace,
\]
and ${\cal B}_{\left[\pm\right]}$ are constants. Note that for the
other copy ($a_{\varphi}=a_{\varphi}^{-}$), the Killing vector-spinors
are obtained from (\ref{eq:vartheta-Sol1}), making $t\rightarrow-t$.

For the allowed class of solutions that fulfills the hypersymmetry
bounds ($m-n=1$), the Killing vector-spinors are characterized by
$\omega{}_{m,n}=\xi-\left(32n^{2}-18n-9\right)\mu/30$ , and $\omega{}_{n,m}=\xi-(32n^{2}+82n+41)\mu/30$.

We also see that the solutions in the non allowed class that violates
the hypersymmetry bounds are nevertheless maximally hypersymmetric.
This result should not come as a surprise because it is very similar
to what is encountered for supersymmetry in $2+1$ supergravity. The
conical surplus solutions with an excess angle that is an integer
multiple of $2\pi$ are maximally supersymmetric since they have the
maximum number of Killing spinors. Nevertheless, they violate the
unitarity bounds that follow from the asymptotic supersymmetry algebra
\cite{Coussaert:1993jp}.

It is of interest to extend the present hypersymmetry analysis to
the Euclidean continuation (``hot solitons'' with inverse temperature
determined by the periodicity in Euclidean time). Remarkably, one
finds that half of the hypersymmetries remain on each copy, provided
the spin-4 chemical potential is suitably switched on. This can be
seen as follows. The Euclidean Killing vector-spinor is given by 
\begin{align}
\vartheta & ={\cal B}_{1}e^{m\left(i\varphi+\omega_{m,n}\tau\right)}+{\cal B}_{2}e^{-m\left(i\varphi+\omega_{m,n}\tau\right)}+{\cal B}_{3}e^{n\left(i\varphi+\omega_{n,m}\tau\right)}+{\cal B}_{4}e^{-n\left(i\varphi+\omega_{n,m}\tau\right)}\thinspace,
\end{align}
where ${\cal B}_{I}$ stand for complex constants. Therefore, since
$\omega_{m,n}\ne\omega_{n,m}$, periodicity in Euclidean time can
only be achieved provided one of the $\omega$'s vanish, and the constants
associated to the other $\omega$ also do. Hence, for the case with
$\omega_{m,n}=0$, the spin-4 chemical potential has to be fixed according
to 
\begin{equation}
\mu=\frac{30}{41m^{2}-9n^{2}}\xi\thinspace,\label{eq:CPFixSo}
\end{equation}
while if $\omega_{n,m}=0$, it is given by (\ref{eq:CPFixSo}) with
$m\leftrightarrow n$.

Note that when the hypersymmetry bounds are fulfilled, eq. (\ref{eq:CPFixSo})
implies that the spin-4 chemical potential has the same sign as $\xi$,
except in the case of ${\cal B}_{3}={\cal B}_{4}=0$, with $n=\frac{1}{2}$
(``hot anti-de Sitter space''), for which $\mu=-3\xi$. It should
be stressed that if the spin-4 chemical potential were absent, all
of the hypersymmetries would be necessarily broken. The existence
of Euclidean solutions with enhanced symmetry might signal a phase
transition. We hope to return to this question in the future.

\section{Concluding remarks \label{sec:Concluding-remarks}}

In this article we have shown that hypersymmetry implies interesting
bounds on the solutions of three-dimensional AdS hypergravity. We
have investigated the black hole solutions of that theory and have
established that these are well defined provided their charges $\mathcal{L}$
and ${\cal U}$ fulfill the inequalities $\mathcal{L}\geq0$, $-\frac{3\pi}{k}{\cal L}^{2}\leq{\cal U}\leq\frac{16\pi}{3k}{\cal L}^{2}$.
We have shown that the inequality $-\frac{3\pi}{k}{\cal L}^{2}\leq{\cal U}$
can be directly inferred from unitarity and the asymptotic $W\left(2,5/2,4\right)$
algebra, which admits a large central charge (classical) limit. Furthermore,
we have proved that the extremal black holes that saturate the hypersymmetry
bound ($-\frac{3\pi}{k}{\cal L}^{2}={\cal U}$) has $\frac{1}{4}$-hypersymmetry.

We have also applied the analysis to soliton-like solutions and have
exhibited the solutions compatible with the hypersymmetry bounds.
We found that for soliton-like solutions, the spin-2 charge fulfills
$\mathcal{L}\leq-\frac{k}{8\pi}$, while the hypersymmetry bound implies
${\cal U}\geq0$. It should be noted in that respect that a different
class of conical defects was discussed in \cite{Chen:2013oxa} in
the case of $Sp\left(4\right)$.

The simplicity of hypergravity is an advantage for bringing hypergravity
bounds in the limelight, but it is at the same time a shortcoming
since one gets only one bound involving the higher spin charge, and
no bound on the asymptotic Virasoro generators. To get more bounds,
and in particular bounds on the asymptotic Virasoro generators from
the asymptotic algebra \cite{Coussaert:1993jp}, one needs to consider
a theory which is not only hypersymmetric, but which possesses also
the traditional supersymmetry with the corresponding generators of
conformal weight $3/2$.

One might expect that a candidate might be the $sl(3\vert2)\oplus sl(3\vert2)$
(more generally, $sl(N\vert N-1)\oplus sl(N\vert N-1)$) higher spin
supergravity theory, which has been studied in \cite{Tan:2011tj,Datta:2012km}.
This theory is a supersymmetric and hypersymmetric extension of the
spin $3$-gravity described by the algebra $sl(3,\mathbb{R})\oplus sl(3,\mathbb{R})$.
The corresponding black holes that manifestly preserve the boundary
conditions have been studied in \cite{HPTT,Bunster:2014mua}. For
some values of the mass, angular momentum and spin-$3$ charges, one
gets extreme black holes, beyond which the entropy ceases to be real.
One might wonder whether these extreme black holes also saturate bounds
following from the asymptotic algebra. However, given the way one
must embed $sl(2,\mathbb{R})$ into $sl(3\vert2)$ in order to avoid
unphysical features (such as fermionic integer spin fields), it turns
out that the spin-$3$ black holes are not solutions of the $sl(3\vert2)\oplus sl(3\vert2)$
theory, unless additional spin-$2$ charges are turned on. This prevents
a discussion of the original (unmodified) bounds of the pure $sl(3,\mathbb{R})$
theory.

Furthermore - and more importantly - there are unitarity difficulties
with the large central charge limit of the corresponding quantum algebra,
preventing an analysis of the type given in the text. It can indeed
be shown that unitary representations of the quantum algebra $W\left(2,3/2,3,5/2\right)$,
also known as $WA_{2}$, exist provided $c$ is suitably bounded from
above \cite{Romans,WReview}.

Finally, we emphasize that we introduced the chemical potentials in
a manner that did not modify the asymptotic symmetry \cite{HPTT}.
Had we proceeded differently, the asymptotic analysis would be different,
and hence, also the bounds that one can derive from it. This is discussed
in Appendix \ref{sec:BHsKRAUS}.

\noindent \acknowledgments We thank A. Campoleoni, O. Fuentealba,
G. Lucena Gómez, J. Matulich and R. Rahman for helpful discussions.
A. P. and R.T. thanks the International Solvay Institutes and the
ULB for warm hospitality. The work of A.P., D.T. and R.T. is partially
funded by the Fondecyt grants N${^{\circ}}$ 11130262, 11130260, 1130658,
1121031. The Centro de Estudios Cient\'{i}ficos (CECs) is funded by
the Chilean Government through the Centers of Excellence Base Financing
Program of Conicyt. The work of M.H. and D.T. is partially supported
by the ERC through the ``SyDuGraM\textquotedblright \ Advanced Grant,
by FNRS-Belgium (convention FRFC PDR T.1025.14 and convention IISN
4.4503.15) and by the ``Communauté Française de Belgique\textquotedblright \ through
the ARC program.

\appendix

\section{Remarks on $osp\left(1|4\right)$ and $WB_{2}$ algebras\label{sec:MR}}

In the fundamental ($5\times5$) matrix representation of $osp\left(1|4\right)$,
the generators can be explicitly written as
\begin{align}
L_{-1} & =\left(\begin{array}{ccccc}
0 & -\sqrt{3} & 0 & 0 & 0\\
0 & 0 & -2 & 0 & 0\\
0 & 0 & 0 & -\sqrt{3} & 0\\
0 & 0 & 0 & 0 & 0\\
0 & 0 & 0 & 0 & 0
\end{array}\right)\;;\;L_{1}=\left(\begin{array}{ccccc}
0 & 0 & 0 & 0 & 0\\
\sqrt{3} & 0 & 0 & 0 & 0\\
0 & 2 & 0 & 0 & 0\\
0 & 0 & \sqrt{3} & 0 & 0\\
0 & 0 & 0 & 0 & 0
\end{array}\right)\;;\;L_{0}=\frac{1}{2}\left(\begin{array}{ccccc}
3 & 0 & 0 & 0 & 0\\
0 & 1 & 0 & 0 & 0\\
0 & 0 & -1 & 0 & 0\\
0 & 0 & 0 & -3 & 0\\
0 & 0 & 0 & 0 & 0
\end{array}\right)\thinspace,\label{eq:-1}
\end{align}
\begin{align}
U_{-3} & =\left(\begin{array}{ccccc}
0 & 0 & 0 & -10 & 0\\
0 & 0 & 0 & 0 & 0\\
0 & 0 & 0 & 0 & 0\\
0 & 0 & 0 & 0 & 0\\
0 & 0 & 0 & 0 & 0
\end{array}\right)\thinspace\thinspace\thinspace;\thinspace\thinspace\thinspace U_{-2}=\frac{5}{\sqrt{3}}\left(\begin{array}{ccccc}
0 & 0 & 1 & 0 & 0\\
0 & 0 & 0 & -1 & 0\\
0 & 0 & 0 & 0 & 0\\
0 & 0 & 0 & 0 & 0\\
0 & 0 & 0 & 0 & 0
\end{array}\right)\thinspace\thinspace\thinspace;\thinspace\thinspace\thinspace U_{-1}=\frac{2}{\sqrt{3}}\left(\begin{array}{ccccc}
0 & -1 & 0 & 0 & 0\\
0 & 0 & \sqrt{3} & 0 & 0\\
0 & 0 & 0 & -1 & 0\\
0 & 0 & 0 & 0 & 0\\
0 & 0 & 0 & 0 & 0
\end{array}\right)\thinspace,\nonumber \\
U_{3} & =\left(\begin{array}{ccccc}
0 & 0 & 0 & 0 & 0\\
0 & 0 & 0 & 0 & 0\\
0 & 0 & 0 & 0 & 0\\
10 & 0 & 0 & 0 & 0\\
0 & 0 & 0 & 0 & 0
\end{array}\right)\thinspace\thinspace\thinspace;\thinspace\thinspace\thinspace U_{2}=\frac{5}{\sqrt{3}}\left(\begin{array}{ccccc}
0 & 0 & 0 & 0 & 0\\
0 & 0 & 0 & 0 & 0\\
1 & 0 & 0 & 0 & 0\\
0 & -1 & 0 & 0 & 0\\
0 & 0 & 0 & 0 & 0
\end{array}\right)\thinspace\thinspace\thinspace;\thinspace\thinspace\thinspace U_{1}=\frac{2}{\sqrt{3}}\left(\begin{array}{ccccc}
0 & 0 & 0 & 0 & 0\\
1 & 0 & 0 & 0 & 0\\
0 & -\sqrt{3} & 0 & 0 & 0\\
0 & 0 & 1 & 0 & 0\\
0 & 0 & 0 & 0 & 0
\end{array}\right)\thinspace,\nonumber \\
U_{0} & =\frac{1}{2}\left(\begin{array}{ccccc}
1 & 0 & 0 & 0 & 0\\
0 & -3 & 0 & 0 & 0\\
0 & 0 & 3 & 0 & 0\\
0 & 0 & 0 & -1 & 0\\
0 & 0 & 0 & 0 & 0
\end{array}\right)\thinspace,
\end{align}
with 
\begin{align}
\mathcal{S}_{-\frac{3}{2}} & =\left(\begin{array}{ccccc}
0 & 0 & 0 & 0 & 1\\
0 & 0 & 0 & 0 & 0\\
0 & 0 & 0 & 0 & 0\\
0 & 0 & 0 & 0 & 0\\
0 & 0 & 0 & -5 & 0
\end{array}\right)\thinspace\thinspace\thinspace;\thinspace\thinspace\thinspace\mathcal{S}_{-\frac{1}{2}}=\frac{1}{\sqrt{3}}\left(\begin{array}{ccccc}
0 & 0 & 0 & 0 & 0\\
0 & 0 & 0 & 0 & 1\\
0 & 0 & 0 & 0 & 0\\
0 & 0 & 0 & 0 & 0\\
0 & 0 & 5 & 0 & 0
\end{array}\right)\thinspace,\nonumber \\
\mathcal{S}_{\frac{1}{2}} & =\frac{1}{\sqrt{3}}\left(\begin{array}{ccccc}
0 & 0 & 0 & 0 & 0\\
0 & 0 & 0 & 0 & 0\\
0 & 0 & 0 & 0 & 1\\
0 & 0 & 0 & 0 & 0\\
0 & -5 & 0 & 0 & 0
\end{array}\right)\thinspace\thinspace\thinspace;\thinspace\thinspace\thinspace\mathcal{S}_{\frac{3}{2}}=\left(\begin{array}{ccccc}
0 & 0 & 0 & 0 & 0\\
0 & 0 & 0 & 0 & 0\\
0 & 0 & 0 & 0 & 0\\
0 & 0 & 0 & 0 & 1\\
5 & 0 & 0 & 0 & 0
\end{array}\right)\thinspace.
\end{align}

In order to carry out the Euclidean continuation, it is worth pointing
out that the $osp\left(4\right)$ matrices fulfill the following properties:
$L_{i}^{\dagger}=\left(-1\right)^{i}L_{-i}$ and $U_{i}^{\dagger}=\left(-1\right)^{i}U_{-i}$.

One easily verifies along standard lines that the asymptotic form
of the $osp\left(1|4\right)$-valued gauge fields $a_{\varphi}^{\pm}$
in (\ref{eq:atheta}) is maintained under gauge transformations given
by 
\begin{equation}
\Omega^{\pm}\left[\epsilon_{\pm},\chi_{\pm},\vartheta_{\pm}\right]=\epsilon_{\pm}\left(t,\varphi\right)L_{\pm1}-\chi_{\pm}\left(t,\varphi\right)U_{\pm3}\mp\vartheta_{\pm}\left(t,\varphi\right)\mathcal{S}_{\pm\frac{3}{2}}+\eta^{\pm}\left[\epsilon_{\pm},\chi_{\pm},\vartheta_{\pm}\right]\thinspace,\label{eq:Omegamn}
\end{equation}
with 
\begin{align}
\eta^{\pm}\left[\epsilon_{\pm},\chi_{\pm},\vartheta_{\pm}\right] & =-\frac{3\pi}{k}\left(i\psi^{\pm}\vartheta_{\pm}+\frac{2}{3}\epsilon_{\pm}\mathcal{L}^{\pm}+2\chi_{\pm}\mathcal{U}^{\pm}-\frac{k}{6\pi}\epsilon_{\pm}^{\prime\prime}\right)L_{\mp1}\mp\epsilon_{\pm}^{\prime}L_{0}\nonumber \\
 & +\frac{6\pi}{k}\left(\chi_{\pm}\mathcal{L}^{\pm}-\frac{k}{12\pi}\chi_{\pm}^{\prime}\right)U_{\pm1}\mp\frac{2\pi}{k}\left(\chi_{\pm}\mathcal{L}^{\pm\prime}+\frac{8}{3}\chi_{\pm}^{\prime}\mathcal{L}^{\pm}-\frac{k}{12\pi}\chi_{\pm}^{\prime\prime\prime}\right)U_{0}\nonumber \\
 & -\frac{\pi}{2k}\left[i\psi^{\pm}\text{\ensuremath{\vartheta}}_{\pm}+2\left(\mathcal{U}^{\pm}-\frac{1}{2}\mathcal{L}^{\pm\prime\prime}+\frac{12\pi}{k}\left(\mathcal{L}^{\pm}\right)^{2}\right)\chi-\frac{11}{3}\chi_{\pm}^{\prime}\mathcal{L}^{\pm\prime}\right.\nonumber \\
 & \left.-\frac{14}{3}\chi_{\pm}^{\prime\prime}\mathcal{L}^{\pm}+\frac{k}{12\pi}\chi_{\pm}^{\left(4\right)}\right]U_{\mp1}\pm\chi_{\pm}^{\prime}U_{\pm2}\pm\frac{\pi}{2k}\left[i\psi^{\pm}\vartheta_{\pm}^{\prime}+\frac{1}{5}i\psi^{\pm\prime}\vartheta_{\pm}\right.\nonumber \\
 & -\frac{5}{3}\chi_{\pm}^{\prime\prime}\mathcal{L}^{\pm\prime}-\frac{4}{3}\mathcal{L}^{\pm}\chi_{\pm}^{\prime\prime\prime}+\frac{2}{5}\left(\mathcal{U}^{\pm}-\frac{1}{2}\mathcal{L}^{\pm\prime\prime}+\frac{18\pi}{k}\left(\mathcal{L}^{\pm}\right)^{2}\right)^{\prime}\chi_{\pm}\nonumber \\
 & \left.+\frac{6}{5}\left(\mathcal{U}^{\pm}-\frac{7}{9}\mathcal{L}^{\pm\prime\prime}+\frac{44\pi}{3k}\left(\mathcal{L}^{\pm}\right)^{2}\right)\chi_{\pm}^{\prime}+\frac{k}{60\pi}\chi_{\pm}^{\left(5\right)}\right]U_{\mp2}\\
 & -\frac{\pi}{4k}\left\{ i\psi^{\pm}\vartheta_{\pm}^{\prime\prime}+\frac{1}{15}i\left(\psi^{\pm\prime\prime}-2^{4}\frac{5\pi}{k}\mathcal{L}^{\pm}\psi^{\pm}\right)\vartheta_{\pm}+\frac{2}{5}i\psi^{\pm\prime}\vartheta_{\pm}^{\prime}\right.\nonumber \\
 & -\chi_{\pm}^{\prime\prime\prime}\mathcal{L}^{\pm\prime}-\frac{4}{5}\epsilon_{\pm}\mathcal{U}^{\pm}+\frac{2}{3}\left(\mathcal{U}^{\pm}-\frac{13}{10}\mathcal{L}^{\pm\prime\prime}+\frac{272\pi}{15k}\left(\mathcal{L}^{\pm}\right)^{2}\right)\chi_{\pm}^{\prime\prime}\nonumber \\
 & +\frac{8}{15}\left(\mathcal{U}^{\pm}-\frac{17}{24}\mathcal{L}^{\pm\prime\prime}+\frac{241\pi}{12k}\left(\mathcal{L}^{\pm}\right)^{2}\right)^{\prime}\chi_{\pm}^{\prime}+\frac{40\pi}{3k}\left[i\psi^{\pm}\psi^{\pm\prime}-\frac{11}{5^{2}}\mathcal{U}^{\pm}\mathcal{L}^{\pm}\right.\nonumber \\
 & \left.-\frac{12\pi}{5k}\left(\mathcal{L}^{\pm}\right)^{3}+\frac{k}{10^{2}\pi}\left(\mathcal{U}^{\pm}-\frac{1}{2}\mathcal{L}^{\pm\prime\prime}\right)^{\prime\prime}+\frac{3^{2}}{5^{2}}\left(\mathcal{L}^{\pm\prime}\right)^{2}+\frac{23}{50}\mathcal{L}^{\pm\prime\prime}\mathcal{L}^{\pm}\right]\chi_{\pm}\nonumber \\
 & \left.-\frac{5}{9}\chi_{\pm}^{\left(4\right)}\mathcal{L}^{\pm}+\frac{k}{180\pi}\chi_{\pm}^{\left(6\right)}\right\} U_{\mp3}-\frac{2\pi}{k}\left[\epsilon_{\pm}\psi^{\pm}+\frac{1}{2}\vartheta_{\pm}\mathcal{L}^{\pm\prime}+\frac{7}{6}\vartheta_{\pm}^{\prime}\mathcal{L}^{\pm}\right.\nonumber \\
 & \left.-\frac{5}{3}\left(\psi^{\pm\prime\prime}+\frac{52\pi}{5k}\mathcal{L}^{\pm}\psi^{\pm}\right)\chi_{\pm}-\frac{25}{6}\chi_{\pm}^{\prime}\psi^{\pm\prime}-\frac{17}{6}\chi_{\pm}^{\prime\prime}\psi^{\pm}-\frac{k}{12\pi}\vartheta_{\pm}^{\prime\prime\prime}\right]{\cal S}_{\mp\frac{3}{2}}\nonumber \\
 & \pm\frac{3\pi}{k}\left(\vartheta_{\pm}\mathcal{L}^{\pm}-\frac{10}{3}\chi_{\pm}\psi^{\pm\prime}-5\chi_{\pm}^{\prime}\psi^{\pm}-\frac{k}{6\pi}\vartheta_{\pm}^{\prime\prime}\right){\cal S}_{\mp\frac{1}{2}}\nonumber \\
 & +\frac{20\pi}{k}\left(\chi_{\pm}\psi^{\pm}+\frac{k}{20\pi}\vartheta_{\pm}^{\prime}\right){\cal S}_{\pm\frac{1}{2}}\thinspace,\nonumber 
\end{align}
provided the fields ${\cal L}^{\pm}$, ${\cal U}^{\pm}$, $\psi^{\pm}$
transform according to
\begin{align}
\mathcal{\delta L}^{\pm} & =2\epsilon_{\pm}^{\prime}\mathcal{L}^{\pm}+\epsilon_{\pm}\mathcal{L}^{\pm\prime}-\frac{k}{4\pi}\epsilon_{\pm}^{\prime\prime\prime}+3\mathcal{U}^{\pm\prime}\chi_{\pm}+4\mathcal{U}^{\pm}\chi_{\pm}^{\prime}+\frac{5}{2}i\psi^{\pm}\vartheta_{\pm}^{\prime}+\frac{3}{2}i\psi^{\pm\prime}\vartheta_{\pm}\thinspace,\nonumber \\
\delta\psi^{\pm} & =\frac{5}{2}\epsilon_{\pm}^{\prime}\psi^{\pm}+\epsilon_{\pm}\psi^{\pm\prime}-\left(\mathcal{U}^{\pm}-\frac{1}{2}\mathcal{L}^{\pm\prime\prime}+\frac{3\pi}{k}\Lambda_{\pm}^{\left(4\right)}\right)\vartheta_{\pm}+\frac{5}{3}\left(\mathcal{L}^{\pm}\vartheta_{\pm}^{\prime}-\frac{k}{20\pi}\vartheta_{\pm}^{\prime\prime\prime}\right)^{\prime}\nonumber \\
 & +\frac{82\pi}{3k}\left(\Lambda_{\pm}^{\left(9/2\right)\prime}-\frac{23}{82}\Lambda_{\text{\ensuremath{\pm}}}^{\left(11/2\right)}-\frac{5k}{82\pi}\psi^{\pm\prime\prime\prime}\right)\chi_{\pm}+\frac{35\pi}{k}\left(\Lambda_{\pm}^{\left(9/2\right)}-\frac{k}{6\pi}\psi^{\pm\prime\prime}\right)\chi_{\pm}^{\prime}\nonumber \\
 & -7\chi_{\pm}^{\prime\prime}\psi^{\pm\prime}-\frac{35}{12}\chi_{\pm}^{\prime\prime\prime}\psi^{\pm}\thinspace,\label{eq:DeltaFields0}
\end{align}
 
\begin{align}
\delta\mathcal{U}^{\pm} & =4\epsilon_{\pm}^{\prime}\mathcal{U}^{\pm}+\epsilon_{\pm}\mathcal{U}^{\pm\prime}+\frac{23\pi}{3k}i\left(\Lambda_{\pm}^{\left(11/2\right)}+\Lambda_{\pm}^{\left(9/2\right)\prime}-\frac{k}{92\pi}\psi^{\pm\prime\prime\prime}\right)\vartheta_{\pm}-\frac{7}{4}i\psi^{\pm\prime}\vartheta_{\pm}^{\prime\prime}\nonumber \\
 & -\frac{35}{12}i\psi^{\pm}\vartheta_{\pm}^{\prime\prime\prime}+\frac{35\pi}{k}i\left(\Lambda_{\pm}^{\left(9/2\right)}-\frac{k}{60\pi}\psi^{\pm\prime\prime}\right)\vartheta_{\pm}^{\prime}-\frac{1}{6}\left[\left(\mathcal{U}^{\pm}-\frac{1}{2}\mathcal{L}^{\pm\prime\prime}\right)^{\prime\prime}\right.\nonumber \\
 & \left.+\frac{144}{k}\left(\Lambda_{\pm}^{\left(6\right)}-\frac{49}{216}\Lambda_{\pm}^{\left(4\right)\prime\prime}\right)\right]^{\prime}\chi_{\pm}-\frac{5}{6}\left[\left(\mathcal{U}^{\pm}-\frac{2}{3}\mathcal{L}^{\pm\prime\prime}\right)^{\prime\prime}+\frac{288}{5k}\Lambda_{\pm}^{\left(6\right)}\right]\chi_{\pm}^{\prime}\nonumber \\
 & +\frac{14}{9}\left(\mathcal{L}^{\pm\prime\prime}-\frac{27}{28}\mathcal{U}^{\pm}-\frac{21\pi}{k}\Lambda_{\pm}^{\left(4\right)}\right)^{\prime}\chi_{\pm}^{\prime\prime}+\frac{7}{3}\left(\mathcal{L}^{\pm\prime\prime}-\frac{3}{7}\mathcal{U}^{\pm}-\frac{28\pi}{3k}\Lambda_{\pm}^{\left(4\right)}\right)\chi^{\prime\prime\prime}\label{eq:DeltaFields}\\
 & +\frac{35}{18}\mathcal{L}^{\pm\prime}\chi_{\pm}^{\left(4\right)}+\frac{7}{9}\mathcal{L}^{\pm}\chi_{\pm}^{\left(5\right)}-\frac{k}{144\pi}\chi_{\pm}^{\left(7\right)}\thinspace,\nonumber 
\end{align}
where 
\begin{align}
\Lambda_{\pm}^{\left(4\right)} & =\left(\mathcal{L}^{\pm}\right)^{2}\thinspace,\nonumber \\
\Lambda_{\pm}^{\left(9/2\right)} & =\mathcal{L}^{\pm}\psi^{\pm}\thinspace,\nonumber \\
\Lambda_{\pm}^{\left(11/2\right)} & =\frac{27}{23}\mathcal{L}^{\pm\prime}\psi^{\pm}\thinspace,\label{eq:NonLinearterms}\\
\Lambda_{\pm}^{\left(6\right)} & =-\frac{7}{18}\mathcal{U}^{\pm}\mathcal{L}^{\pm}-\frac{8\pi}{3k}\left(\mathcal{L}^{\pm}\right)^{3}+\frac{295}{432}\left(\mathcal{L}^{\pm\prime}\right)^{2}+\frac{22}{27}\mathcal{L}^{\pm\prime\prime}\mathcal{L}^{\pm}+\frac{25}{12}i\psi^{\pm}\psi^{\pm\prime}\thinspace.\nonumber 
\end{align}
Here prime denotes the derivative with respect to $\varphi$. Note
that the appearance of the imaginary unit ``$i$'' in the product
of (real) Grassmann variables is because we have adopted the following
convention: $\left(\text{\ensuremath{\vartheta}}_{1}\text{\ensuremath{\vartheta}}_{2}\right)^{*}=\text{\ensuremath{\vartheta}}_{2}^{*}\text{\ensuremath{\vartheta}}_{1}^{*}=-\text{\ensuremath{\vartheta}}_{1}\text{\ensuremath{\vartheta}}_{2}$.

The transformation law of the fields then allows one to readily find
the Poisson bracket algebra of the canonical generators, which for
each copy reads
\begin{align}
\left[{\cal L}\left(\varphi\right),{\cal L}\left(\phi\right)\right]_{PB} & =-2\delta^{\prime}\left(\varphi-\phi\right)\mathcal{L}\left(\varphi\right)-\delta\left(\varphi-\phi\right)\mathcal{L}^{\prime}\left(\varphi\right)+\frac{k}{4\pi}\delta'''\left(\varphi-\phi\right)\thinspace,\nonumber \\
\left[{\cal L}\left(\varphi\right),{\cal U}\left(\phi\right)\right]_{PB} & =-4\delta^{\prime}\left(\varphi-\phi\right)\mathcal{U}\left(\varphi\right)-3\delta\left(\varphi-\phi\right)\mathcal{U}^{\prime}\left(\varphi\right)\thinspace,\nonumber \\
\left[{\cal L}\left(\varphi\right),\psi\left(\phi\right)\right]_{PB} & =-\frac{5}{2}\delta^{\prime}\left(\varphi-\phi\right)\psi\left(\varphi\right)-\frac{3}{2}\delta\left(\varphi-\phi\right)\psi^{\prime}\left(\varphi\right)\thinspace,\nonumber \\
\left[{\cal U}\left(\varphi\right),\psi\left(\phi\right)\right]_{PB} & =\frac{7}{12}\left(\psi^{\prime\prime}\left(\varphi\right)-\frac{60\pi}{k}\Lambda^{\left(9/2\right)}\left(\varphi\right)\right)\delta^{\prime}\left(\varphi-\phi\right)+\frac{7}{4}\psi^{\prime}\delta^{\prime\prime}\left(\varphi-\phi\right)\nonumber \\
 & +\frac{1}{12}\left[\psi'''\left(\varphi\right)-\frac{92\pi}{k}\left(\Lambda^{\left(11/2\right)}\left(\varphi\right)+\Lambda^{\left(9/2\right)\prime}\left(\varphi\right)\right)\right]\delta\left(\varphi-\phi\right)\nonumber \\
 & +\frac{35}{12}\psi\left(\varphi\right)\delta'''\left(\varphi-\phi\right)\thinspace,\label{eq:}
\end{align}
 
\begin{align}
\left[{\cal U}\left(\varphi\right),{\cal U}\left(\phi\right)\right]_{PB} & =\frac{5}{6}\left[\left(\mathcal{U}\left(\varphi\right)-\frac{2}{3}\mathcal{L}^{\prime\prime}\left(\varphi\right)\right)^{\prime\prime}+\frac{288\pi}{5k}\Lambda^{\left(6\right)}\left(\varphi\right)\right]\delta^{\prime}\left(\varphi-\phi\right)\nonumber \\
 & +\frac{1}{6}\left[\left(\mathcal{U}\left(\varphi\right)-\frac{1}{2}\mathcal{L}^{\prime\prime}\left(\varphi\right)-\frac{98\pi}{3k}\Lambda^{\left(4\right)}\left(\varphi\right)\right)^{\prime\prime}+\frac{144\pi}{k}\Lambda^{\left(6\right)}\left(\varphi\right)\right]^{\prime}\delta\left(\varphi-\phi\right)\nonumber \\
 & +\frac{3}{2}\left(\mathcal{U}\left(\varphi\right)-\frac{28}{27}\mathcal{L}^{\prime\prime}\left(\varphi\right)+\frac{196\pi}{9k}\Lambda^{\left(4\right)}\left(\varphi\right)\right)^{\prime}\delta^{\prime\prime}\left(\varphi-\phi\right)\\
 & +\left(\mathcal{U}\left(\varphi\right)-\frac{7}{3}\mathcal{L}^{\prime\prime}\left(\varphi\right)+\frac{196\pi}{9k}\Lambda^{\left(4\right)}\left(\varphi\right)\right)\delta'''\left(\varphi-\phi\right)\nonumber \\
 & -\frac{35}{18}\mathcal{L}^{\prime}\left(\varphi\right)\delta^{\left(4\right)}\left(\varphi-\phi\right)-\frac{7}{9}\mathcal{L}\left(\varphi\right)\delta^{\left(5\right)}\left(\varphi-\phi\right)+\frac{k}{144\pi}\delta^{\left(7\right)}\left(\varphi-\phi\right)\thinspace,\nonumber \\
i\left[\psi\left(\varphi\right),\psi\left(\phi\right)\right]_{PB} & =\left(\mathcal{U}\left(\varphi\right)-\frac{1}{2}\mathcal{L}^{\prime\prime}\left(\varphi\right)+\frac{3\pi}{k}\Lambda^{\left(4\right)}\left(\varphi\right)\right)\delta(\varphi-\phi)-\frac{5}{3}\mathcal{L}^{\text{\ensuremath{\prime}}}\left(\varphi\right)\delta^{\prime}\left(\varphi-\phi\right)\nonumber \\
 & -\frac{5}{3}\mathcal{L}\left(\varphi\right)\delta^{\prime\prime}\left(\varphi-\phi\right)+\frac{k}{12\pi}\delta^{\left(4\right)}\left(\varphi-\phi\right)\thinspace,\nonumber 
\end{align}
(the Poisson bracket between two fermionic fields being symmetric).

\section{Black holes in a different embedding \label{sec:BHsKRAUS}}

In the case of an action principle described by two copies of $sp\left(4\right)$,
a different proposal for black hole solutions has been given in \cite{Chen-Long-Wang-BH}.
That article follows the lines of \cite{GK}, \cite{AGKP}, and hence
the solution does not fit within our boundary conditions. For the
plus copy of the gauge field, which we denote by $\tilde{a}^{+}$,
the solution reads\footnote{The relationship between the variables in (\ref{eq:CLW-BHs}) and
the ones in \cite{Chen-Long-Wang-BH}, here denoted by ${\cal \tilde{L}}$,
${\cal \tilde{W}}$, $\tilde{\mu}$, and $W_{n}^{\left(4\right)}$,
is given by ${\cal \tilde{L}}=\frac{2\pi}{k}\mathcal{\hat{L}}\;,\;{\cal \tilde{W}}=\frac{\pi}{3k}\mathcal{\hat{U}}\;,\;\tilde{\mu}=-\frac{5}{3}\hat{\mu}\thinspace,$$W_{n}^{\left(4\right)}=\frac{3}{5}U_{n}\thinspace.$ } \textit{ 
\begin{align}
\tilde{a}^{+} & =\left(L_{1}-\frac{2\pi}{k}\mathcal{\hat{L}}^{+}L_{-1}+\frac{\pi}{5k}\mathcal{\hat{U}}^{+}U_{-3}\right)dx^{+}+\hat{\mu}^{+}\left[-U_{3}+\frac{6\pi}{k}\mathcal{\hat{L}}^{+}U_{1}-\frac{6\pi}{k}\mathcal{\hat{U}}^{+}L_{-1}\right.\nonumber \\
 & \left.+\frac{22\pi^{2}}{15k^{2}}\left(\mathcal{\hat{U}}^{+}+\frac{60\pi}{11k}\mathcal{\hat{L}}^{+}\right)\mathcal{\hat{L}}^{+}U_{-3}-\frac{\pi}{k}\left(\mathcal{\hat{U}}^{+}+\frac{12\pi}{k}\left(\mathcal{\hat{L}}^{+}\right)^{2}\right)U_{-1}\right]dx^{-}\thinspace,\label{eq:CLW-BHs}
\end{align}
}where $x^{\pm}=\hat{t}/\ell\pm\varphi$.

The asymptotic form of $\tilde{a}_{\varphi}^{+}$ does not match the
one of our boundary conditions (\ref{eq:atheta}), and exactly as
shown in \cite{Bunster:2014mua}, it can be proved that after a suitable
permissible gauge transformation, the gauge field can be written in
terms of highest weights in the non-principal embedding that corresponds
to the decomposition: $sp\left(4\right)={\cal D}_{1}\oplus2{\cal D}_{1/2}\oplus3{\cal D}_{0}$.
This solution cannot carry therefore spin-4 charges, but instead,
apart from the mass and the angular momentum, it is endowed with two
bosonic spin-3/2 and three spin-1 charges per copy.

This class of solutions are described by the Chern-Simons action principle
based on two copies of $osp\left(1|4\right)$, as in the text, but
with a different set of boundary conditions that are adapted to the
highest weight generators of the alternative embedding. The decomposition
of the full $osp\left(1|4\right)$ is now given by $osp\left(1|4\right)=\left({\cal D}_{1}\oplus2{\cal D}_{1/2}\oplus3{\cal D}_{0}\right)\oplus\left({\cal D}_{1/2}\oplus2{\cal D}_{0}\right)$,
so that it includes also one Majorana gravitino, and two Grassmann-valued
spin-1 fields, per copy.

\medskip{}

\textit{Black hole thermodynamics: }In order to match our conventions,
the Euclidean continuation of (\ref{eq:CLW-BHs}) is obtained through
$\hat{t}=-\ell\hat{\xi}i\tau$, with $0<\tau\leq1$. Hence, in the
branch that is continuously connected with the BTZ black hole, the
regularity condition of the solution, given by (\ref{eq:Holo-tau}),
implies that the chemical potentials turn out to be fixed according
to 
\begin{align}
\hat{\xi} & =\frac{\pi}{50}\left[\frac{\left(3\hat{\lambda}_{\left[+\right]}+\hat{\lambda}_{\left[-\right]}\right)\left(3\left(\hat{\lambda}_{\left[+\right]}^{2}-3^{2}\hat{\lambda}_{\left[-\right]}^{2}\right)+40\hat{\lambda}_{\left[+\right]}\hat{\lambda}_{\left[-\right]}\right)}{\left(\hat{\lambda}_{\left[-\right]}^{2}-\hat{\lambda}_{\left[+\right]}^{2}\right)\hat{\lambda}_{\left[-\right]}\hat{\lambda}_{\left[+\right]}}\right]\thinspace,\\
\hat{\mu} & =\frac{30\left(\hat{\lambda}_{\left[+\right]}-3\hat{\lambda}_{\left[-\right]}\right)}{\left(3\hat{\lambda}_{\left[+\right]}+\hat{\lambda}_{\left[-\right]}\right)\left(3\left(\hat{\lambda}_{\left[+\right]}^{2}-3^{2}\hat{\lambda}_{\left[-\right]}^{2}\right)+40\hat{\lambda}_{\left[+\right]}\hat{\lambda}_{\left[-\right]}\right)}\thinspace,
\end{align}
where 
\begin{equation}
\hat{\lambda}_{\left[\pm\right]}^{2}=\frac{10\pi}{k}\left(\hat{\mathcal{L}}\pm\frac{4}{5}\sqrt{\mathcal{\hat{L}}^{2}-\frac{3k}{16\pi}\hat{\mathcal{U}}}\right)\thinspace,
\end{equation}
stand for the eigenvalues of $\tilde{a}_{+}^{+}$. Thus, the corresponding
Euclidean entropy (\ref{eq:entropy}) is given by 
\begin{equation}
S=\frac{2\pi k}{5}\mbox{Re}\left(3\lambda_{\left[+\right]}+\lambda_{\left[-\right]}\right)=\frac{4\pi k}{5}\text{Re}\left[\frac{\left(\hat{\lambda}_{\left[+\right]}^{2}+\hat{\lambda}_{\left[-\right]}^{2}\right)\left(3^{3}\left(\hat{\lambda}_{\left[+\right]}^{2}-\hat{\lambda}_{\left[-\right]}^{2}\right)+2^{7}\hat{\lambda}_{\left[+\right]}\hat{\lambda}_{\left[-\right]}\right)}{9\left(\hat{\lambda}_{\left[+\right]}^{3}-3\hat{\lambda}_{\left[-\right]}^{3}\right)+41\left(3\hat{\lambda}_{\left[+\right]}-\hat{\lambda}_{\left[-\right]}\right)\hat{\lambda}_{\left[+\right]}\hat{\lambda}_{\left[-\right]}}\right]\thinspace,\label{eq:entropy_diagonal}
\end{equation}
which disagrees with the expression found in \cite{Chen-Long-Wang-BH},
that reads $\hat{S}=\frac{2\pi k}{5}\mbox{Re}\left(3\hat{\lambda}_{\left[+\right]}+\hat{\lambda}_{\left[-\right]}\right)$.
Nonetheless, the bounds that guarantee that the entropy is a real
nonnegative quantity, saturate for the same cases in both expressions,
which corresponds to $\hat{\lambda}_{\left[-\right]}=0$ ($\hat{\mathcal{U}}=-\frac{3\pi}{k}\hat{\mathcal{L}}^{2}$),
and $\hat{\lambda}_{\left[+\right]}=\hat{\lambda}_{\left[-\right]}=0$
($\hat{\mathcal{U}}=\hat{\mathcal{L}}=0$), so that the holonomy around
the thermal circle is no longer trivial. In these cases, the Killing
vector-spinor equation (\ref{eq:KVS-eq}) admits a single globally
defined solution, described by a constant Killing vector-spinor given
by (\ref{eq:theta}), with $\vartheta=\vartheta_{0}$ and $\mathcal{L}=\hat{\mathcal{L}}$. 

{} }
\end{document}